\documentclass[iop, numberedappendix]{emulateapj}
\usepackage{grffile}  
\usepackage{amssymb}
\usepackage{amsmath}
\usepackage{graphicx}
\usepackage{timing}
\usepackage{ISM}
\usepackage{color}
\usepackage{bm}
\usepackage{xspace}
\usepackage{hyperref}

\newcommand{\be}{\begin{eqnarray}}
\newcommand{\ee}{\end{eqnarray}}

\def\gtrsim{\mathrel{\hbox{\rlap{\hbox{\lower4pt\hbox{$\sim$}}}\hbox{$>$}}}}
\def\lesssim{\mathrel{\hbox{\rlap{\hbox{\lower4pt\hbox{$\sim$}}}\hbox{$<$}}}}

\newcommand{\xvecbar}{{\bf \overline{x}}}

\newcommand{\rvec}{{\bf r}}

\newcommand{\qvec}{{\bf q}}

\newcommand{\vvec}{{\bf v}}
\newcommand{\xpsrvec}{{{\bf x}_{p}}}
\newcommand{\xevec}{{{\bf x}_{e}}}
\newcommand{\vpsrvec}{{{\bf v}_p}}
\newcommand{\vevec}{{{\bf v}_e}}

\newcommand{\nhat}{{\hat {\bf n}}}

\newcommand{\zet}{z_{e}}
\newcommand{\zpt}{z_{p}}
\newcommand{\zei}{z_{e_0}}
\newcommand{\zpi}{z_{p_0}}
\newcommand{\zvec}{{\bf z}}
\newcommand{\zhat}{\hat\zvec}
\newcommand{\vepar}{{v_{e_\parallel}}}
\newcommand{\vppar}{{v_{p_\parallel}}}
\newcommand{\veperp}{{\vvec_{e_\perp}}}
\newcommand{\vpperp}{{\vvec_{p_\perp}}}
\newcommand{\dvperp}{\vpperp - \veperp}
\newcommand{\Pne}{P_{\delta n_e}}
\newcommand{\nebar}{{\overline{n}_e}}

\renewcommand{\DM}{\ensuremath{\mathrm{DM}}}
\renewcommand{\veff}{{v_{\rm eff}}}
\renewcommand{\veffperp}{{v_{\rm eff_{\perp}}}}
\renewcommand{\veffperpvec}{{\bf v_{\rm eff_{\perp}}}}	
\renewcommand{\cnsq}{{\rm C_n^2}}
\newcommand{\Dnuiss}{\Delta\nu_{\rm ISS}}
\newcommand{\tauiss}{\tau_{\rm ISS}}

\newcommand{\DMsw}{\ensuremath{\mathrm{DM_{sw}}}}
\newcommand{\DMhel}{\ensuremath{\mathrm{DM_{hel}}}}

\newcommand{\AU}{\ensuremath{\mathrm{AU}}}

\newcommand{\cmthree}{\ensuremath{\mathrm{cm^{-3}}}\xspace}
\newcommand{\apar}{{a_{\parallel}}}
\newcommand{\aperp}{{a_{\perp}}}

\newcommand{\Nec}{{N_{e_c}}}
\newcommand{\nec}{{n_{e_c}}}
\newcommand{\dtbary}{{\Delta t_{\rm bary}}}
\newcommand{\dtgeo}{{\Delta t_{\rm geo}}}

\newcommand{\nuGHz}{\nu_{\rm GHz}}

\shorttitle{Pulsar Dispersion Measure Variations}
\shortauthors{Lam et al.}

\begin{document}

\title{Systematic and Stochastic Variations in Pulsar Dispersion Measures}
\author{
M.\,T.\,Lam\altaffilmark{1},
J.\,M.\,Cordes\altaffilmark{1},
S.\,Chatterjee\altaffilmark{1},
M.\,L.\,Jones\altaffilmark{2},
M.\,A.\,McLaughlin\altaffilmark{2},
J.\,W.\,Armstrong\altaffilmark{3}
}
\altaffiltext{1}{Department of Astronomy and Cornell Center for Astrophysics and Planetary Science, Cornell University, Ithaca, NY 14853; mlam@astro.cornell.edu}
\altaffiltext{2}{Department of Physics and Astronomy, West Virginia University, Morgantown, WV 26506, USA}
\altaffiltext{3}{Jet Propulsion Laboratory, California Institute of Technology, 4800 Oak Grove Drive, Pasadena, CA 91109, USA}

\begin{abstract}
We analyze deterministic and random temporal variations in dispersion measure (DM) from the full three-dimensional velocities of pulsars with respect to the solar system, combined with electron-density variations on a wide range of length scales. Previous treatments have largely ignored the pulsar's changing distance while favoring interpretations involving the change in sky position from transverse motion. Linear trends in pulsar DMs seen over 5-10~year timescales may signify sizable DM gradients in the interstellar medium (ISM) sampled by the changing direction of the line of sight to the pulsar. We show that motions parallel to the line of sight can also account for linear trends, for the apparent excess of DM variance over that extrapolated from scintillation measurements, and for the apparent non-Kolmogorov scalings of DM structure functions inferred in some cases. Pulsar motions through atomic gas may produce bow-shock ionized gas that also contributes to DM variations. We discuss possible causes of periodic or quasi-periodic changes in DM, including seasonal changes in the ionosphere, annual variation of the solar elongation angle, structure in the heliosphere-ISM boundary, and substructure in the ISM. We assess the solar cycle's role on the amplitude of ionospheric and solar-wind variations. Interstellar refraction can produce cyclic timing variations from the error in transforming arrival times to the solar system barycenter. We apply our methods to DM time series and DM gradient measurements in the literature and assess consistency with a Kolmogorov medium. Finally, we discuss the implications of DM modeling in precision pulsar timing experiments.
\end{abstract}


\section{Introduction}

Free electrons in the interstellar medium (ISM) affect pulsar signals by introducing a frequency-dependent dispersion delay. Dispersion delays need to be removed as part of search algorithms in pulsar surveys and for precision time-of-arrival (TOA) measurements that are used for determinations of orbital elements, tests of General Relativity and other theories of gravity, and detection of long-wavelength gravitational waves. Besides being used for correction, dispersion measures (DMs) are the the primary means for determining electron column densities on Galactic and, in some cases, extragalactic lines of sight (LOSs). They serve as important input data for Galactic models of the electron density and in studies of stochastic variations in electron density on length scales $\sim$ 1--100~AU. Dispersion and scattering, a related frequency-dependent phenomenon due to multi-path propagation, are assumed to result from cold plasma in the high-frequency limit with negligible contributions from magnetic fields (see \citealt{Cordes2002} or \citealt{handbook} for a review).

In this paper, we discuss the inferences that can be made about the ISM using epoch-dependent DM measurements.  We analyze DMs  in terms of the full three-dimensional motions of pulsars, the changes in electron density along the entire LOS, and the solar system motion through the ISM. The dispersion measure is the LOS  integral
\be
\DM(t) = \int_0^{D(t)} \!\!\!\!\! ds\, n_e(s \nhat(t), t),
\ee
where  $D(t)$ is the  pulsar's distance, $n_e$ is the electron density, and $\nhat(t)$ is a unit vector from the observer to the pulsar, with all three quantities generally epoch-dependent.   Many pulsars  have much higher velocities than bulk ISM motion, so variations in $\DM$ are usually dominated by the changing LOS, including both the distance and direction. Therefore, we generally drop the explicit time dependence of the electron density, though we will show that this assumption does not hold within the solar system.  While epoch-dependent distances are an obvious consequence of high velocities, most quantitative analyses of DMs have focused on how the LOS changes from transverse motions.

We report on measured DM variations in the literature in \S\ref{sec:measuredDM}. In \S\ref{sec:LOSintegrals} we develop the formalism for DM variations from changing LOS integrals through electron-filled media and we discuss resulting linear trends in DM time series in \S\ref{sec:linear_trends}. We consider the DM struture function (SF) and contributions to it from stochastic DM variations in \S\ref{sec:sf_and_spectrum}. In \S\ref{sec:refraction_effects} we discuss the impact of refraction on timing delays and subsequently the measured DM. We interpret the causes of linear and non-monotonic trends seen in several pulsars in the literature in \S\ref{sec:interpretation_phenomena} and periodic DM variations in \S\ref{sec:annual_trends}. In \S\ref{sec:implications}, we report the impact of DM variations on ISM study and on timing precision. We summarize our findings and conclusions in \S\ref{sec:conclusions}. A list of symbols and acronyms used throughout the paper can be found in Table~\ref{table1}.

\section{Measured DM Variations}
\label{sec:measuredDM}

Time variability in DM is a well-known phenomenon in pulsar timing. Epoch-dependent variations were first detected in the Crab Pulsar \citep{rr1971}. \citet{ir1977} measured DM variations in the Crab Pulsar over a five-year span and suggest that the changes in DM come from within the Crab Nebula. \citet{hhc1985} found a large gradient in the DM of the Vela Pulsar over 15 years and attributed it to the LOS changing with the transverse velocity of the pulsar relative to the supernova remnant. Spatial variations in DM on sub-parsec scales have also been seen (see \citealt{mlr1991} and \citealt{fcl2001}, who discuss changes in DM over different LOSs to pulsars in the center of the globular cluster 47 Tucanae).

Published time series of DM in the literature show several types of behavior. Some show deterministic linear trends superposed with correlated, stochastic variations.  A few show piecewise linear variations that signify  change points in the time derivative $d\DM/dt$ associated with structure in the ISM on scales of 1--100~AU. Many also show periodic variations, either smoothly sinusoidal or sharp with distinct features, often with a period of roughly one year. The amplitudes of these variations have also been seen to change with time. In some cases, both linear and periodic variations are seen. Others show only correlated, stochastic variations without an obvious trend.

\citet{1991ApJ...382L..27P} reported results on five pulsars, four of which show long-term trends  with slopes $\vert d\DM/dt\vert \sim 10^{-3}$~pc~\cmthree~yr$^{-1}$ (increasing: PSRs B0823+26, B0834+06, and B1237+25; decreasing: PSR B0919+06). They assert that the rms of the DM variations is correlated with the average \DM\, but with significant scatter about a best fit relation  $\sigma_{\DM} \propto \DM^{1.3\pm 0.3}$. A trend of this type would generally signify that the DM variations are associated with accumulated effects along the LOS, but the correlation is affected by the long-term trends that may be due to  parallel motion through ionized gas near the pulsars.  

\citet[][see also \citealt{y+07, 2013MNRAS.435.1610P}]{Keith+2013} give $\DM(t)$ time series over roughly $\sim 6$~yr for 20 millisecond pulsars (MSPs) that are monitored in the  Parkes Pulsar Timing Array (PPTA) program. Of these, 11 show  prevailing trends of  increasing DM (PSRs J1024$-$0719, J1730$-$2304, J1732$-$5049, and J1857+0943)  or decreasing DM  (PSRs J1045$-$4509, J1600$-$3053, J1643$-$1224, J1744$-$1134, J1909$-$3744, J1939+2134 and J2129$-$5721).  Two others show overall trends but with a localized DM `event' that breaks the trend  (PSRs J1603$-$7202 and J1824$-$2452).  The remaining seven objects show  non-monotonic variations with various degrees  and timescales of temporal correlation. \citet{Reardon+2016} find evidence for significant linear trends in 13 pulsars and sinusoidal, annual variations in four pulsars in the extended PPTA data release 1. The approximate derivative for PSR J1939+2134 (B1937+21) is about half the value of the 20-year trend reported by  \citet{Ramachandran+2006} and is consistent with changes in slope seen in the 20-year time series. 
  
\citet[][]{Demorest+2013} present $\DM(t)$ time series for 14 out of 17 pulsars that were part of the first data release of the North American Nanohertz Observatory for Gravitational Waves (NANOGrav), based on five years of data.  Of these,  the seven objects that overlap with the \citet{Keith+2013} sample show consistent trends. Of the others, several  objects show very weak DM variation while two pulsars, PSRs B1855+09 and J2317+1439, show strong trends superposed with correlated, stochastic variations.

\begin{figure}[t!]
\hspace{-5ex}
\includegraphics[width=0.52\textwidth]{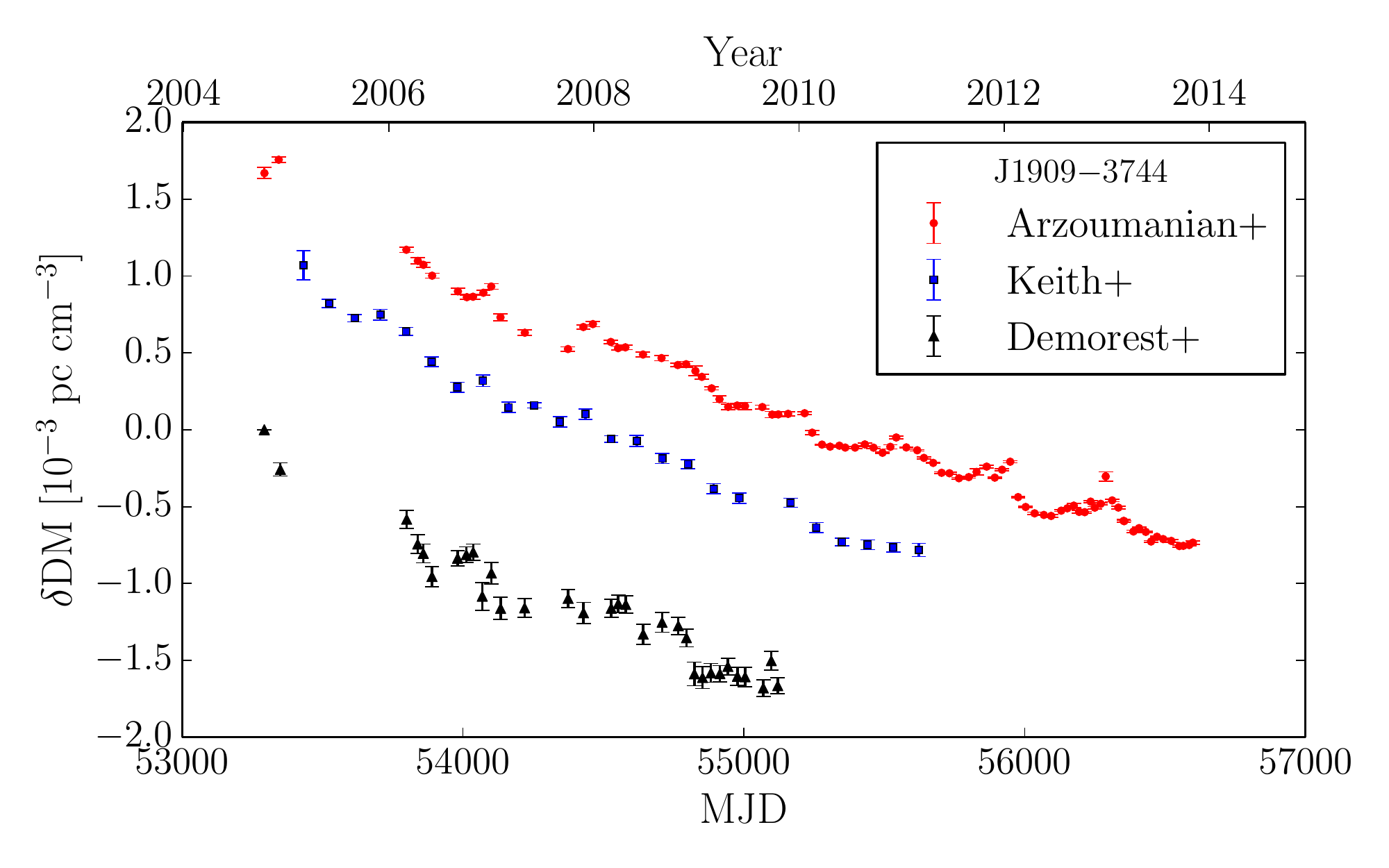}
\caption{\footnotesize DM offsets $\delta \DM(t) = \DM(t) - \DM_{\mathrm{nominal}}$ for PSR J1909-3744, reported in \citet[][red circles]{Arzoumanian+2015}, \citet[][blue squares]{Keith+2013}, \citet[][black triangles]{Demorest+2013}. $\DM_{\mathrm{nominal}} = 10.394680, 10.392717, 10.392031$~pc~\cmthree, respectively. The nominal DM differs due to different methods to account for frequency-dependent pulse shape changes in the timing models.
\label{fig:smallplot}
}
\hspace{3ex}
\end{figure}

\begin{deluxetable*}{llc}
\centering
\tablecolumns{3}
\tablecaption{Symbols and Acronyms Used}
\tablehead{
\colhead{Symbols} & \colhead{Definition} & \colhead{Characteristic Units}\\
}
\startdata
$a$ & Characteristic scale of ISM structure & Length\\
$A$ & Spectral coefficient & \\
$c$ & Speed of light & cm s$^{-1}$\\
$C$ & Arbitrary amplitude & \\
$C_1$ & Constant in uncertainty relation, $2\pi\Dnuiss\tauiss = C_1$ & \\
$\cnsq$ & Coefficient in electron-density wavenumber spectrum & Length$^{-(\beta+3)}$\\
$D$ & Earth-Pulsar distance & kpc\\
DM & Dispersion Measure & pc \cmthree\\
$D_\DM$ & DM structure function & $\mathrm{[pc\;cm^{-3}]^2}$\\
$D_t$ & Time structure function & s$^2$\\
$D_\phi$ & Phase structure function & $\mathrm{radians^2}$\\
$\dot{E}$ & Pulsar energy loss rate & erg s$^{-1}$\\
$f$ & Spectral frequency & Time$^{-1}$\\
$f_1$,$f_2$ & Lower and upper spectral cutoffs of $S_{\DM}$ & Time$^{-1}$\\
$f_\beta$ & Numerical factor in DM structure function &\\
$h$ & Planck constant & erg s\\
$h_s$ & Height above Earth's surface & km\\
$H$ & Characteristic thickness of ionospheric layer & km\\
$k$ & Boltzmann constant  & erg K$^{-1}$ \\
$K$ & Dispersion constant ($\equiv c r_e / 2\pi$) & ms GHz$^2$ pc$^{-1}$ \cmthree\\
$l$ & Characteristic scale of ISM structure & Length\\
$l_g$,$b_g$ & Galactic coordinates (longitude, latitude) & deg\\
$l_{\rm HI}$ & Mean free path for neutral-hydrogen-ionizing radiation & cm\\
$m_p$ & Proton mass & g\\
$n_e$ & Electron density & \cmthree\\
$n_{\rm HI}$ & Effective hydrogen density & \cmthree\\
$n_p$ & Proton density & \cmthree\\
$N_e$ & Electron column density & cm$^{-2}$\\
$N_{\rm HI}$ & Hydrogen column density & cm$^{-2}$\\
$\Pne$ & Wavenumber spectrum for the electron density & Length$^{-3}$\\
$q$ & Wavenumber & Length$^{-1}$\\
$q_1$,$q_2$ & Lower and upper wavenumber cutoffs of $\Pne$ & Length$^{-1}$\\
$r$ & Position & Length\\
$r_e$ & Classical electron radius & cm\\
$r_s$ & Bow-shock standoff radius & cm\\
$r_\oplus$ & Earth-Sun distance & AU\\
$R_{d\DM/dt}$ & ratio of linear trend to rms linear trend from a Kolmogorov medium& \\
$R_{\rm rms}$ & ratio of rms DM after and before a linear trend is removed & \\
$s$ & Represents a generic position along the LOS & Length\\
$S$ & Power spectrum & \\
SM & Scattering Measure & kpc m$^{-(\beta+3)}$\\
$t$ & Time & Time \\
$T$ & Total observing span & Time \\
$v$ & Velocity & km/s \\
$x$ & Position & Length\\
$z$ & Represents a position along the LOS & Length\\
$\alpha$ & Arbitrary spectral index & \\
$\alpha_e$,$\delta_e$ & Equatorial coordinates (RA, declination) & deg\\
$\beta$ & Exponent in wavenumber spectrum for $n_e$ & \\
$\Gamma$ & Gamma function & \\
$\gamma$ & Exponent in power-law of red noise process & \\
$\Delta$ & Difference/Increment & \\
$\Dnuiss$ & Scintillation bandwidth & MHz\\
$\Delta t$ & Time delay & Time\\
$\Dtiss$ & Scintillation timescale & s\\
$\eta_s$ & Shock compression factor & \\
$\theta_i$ & Incidence angle & rad\\
$\theta_r$ & Refraction angle & rad\\
$\theta_z$ & Zenith angle & deg\\
$\lambda, \nu$ & Electromagnetic wavelength and frequency & cm, GHz\\
$\lambda_e$,$\beta_e$ & Ecliptic coordinates (longitude, latitude) & deg\\
$\lambda_h$,$\beta_h$ & Heliographic coordinates (longitude, latitude) & deg\\
$\rho$ & Mass density & g \cmthree \\
$\sigma$ & rms & \\
$\sigma_{\rm HI}$ & Photoionoization cross section for neutral hydrogen & cm$^{-2}$\\
$\tau$ & Time lag & Time\\
$\tauiss$ & Scattering timescale & $\mu$s\\
$\phi$ & Phase perturbation from refractive index perturbations & rad\\
$\phi_g$,$\lambda_g$ & Geographic coordinates (latitude, longitude) & deg\\
$\varphi$ & Sinusoidal phase & rad\\
$\omega$ & Sinusoidal angular frequency & Angle/Time
\enddata
\label{table1}
\end{deluxetable*}

PSR J1909$-$3744 exemplifies several types of variations in DM that motivate our study. \citet{Demorest+2013} see a monotonic decrease in DM over 5 years. \citet{Keith+2013} also note the linearity of $\DM(t)$, with a change in $1.85\times 10^{-3}$~pc~\cmthree over 6 years, and find that the SF of their time series exceeds, for every lag, the SF prediction from dynamic spectrum estimates by a factor of $\sim 5$. They suggest the SF excess implies an electron-density wavenumber spectrum steeper than that of a turbulent, Kolmogorov medium. Recently, the NANOGrav Nine-Year Data Release (\citealt{Arzoumanian+2015}; hereafter NG9) showed that the decreasing trend continued, spanning all nine years of data, along with a superposed annual variation. Figure~\ref{fig:smallplot} shows the DM offsets $\delta \DM(t) = \DM(t) - \DM_{\mathrm{nominal}}$ as presented by the three data sets, where $\DM_{\mathrm{nominal}} = 10.394680$~pc~\cmthree for \citet{Demorest+2013}, 10.392717~pc~\cmthree for \cite{Keith+2013}, and 10.392031~pc~\cmthree for NG9. Differences in the absolute DM are caused by different methods of frequency-dependent pulse shape variation removal from the TOAs. An in-depth analysis of the DM variations of all of the MSPs in NG9 will be presented in the future (M.\ L.\ Jones et al.\ in preparation).

\citet{fst2014} present $\DM(t)$ for the relativistic binary PSR B1534+12 and fit for derivatives $d\DM/dt$ in five separate time blocks. The overall trend is a decrease with time that is interrupted by episodic flattenings or increases in DM.  The variation of $\DM(t)$ from 1990 to 2012 is dominated by five piecewise linear segments lasting three to five years with slopes $d\DM/dt = \{-3.16, -0.43, -2.94, 10.1, -0.1\} \times 10^{-4}$~pc~\cmthree~yr$^{-1}$. The DM SF scales as $\tau^{3.70\pm0.04}$ for lags between 70 and 90 days, consistent with a Kolmogorov scaling. The best-fit SF implies a diffractive scintillation timescale of $\Dtiss = 3.0\pm 0.8$~min at 0.43~GHz, considerably smaller than the range $11\pm3$~min directly measured by \citet{Bogdanov+2002} from 2D autocorrelation functions of dynamic spectra. While epoch-dependent scintillation may play a role in this difference, the shorter time scale inferred from the SF fit is consistent with the presence of contamination from non-Kolmogorov fluctuations on length scales relevant to the DM variations.

\section{Line of Sight Integrals}
\label{sec:LOSintegrals}

\begin{figure}[t!]
\hspace{10ex}
\includegraphics[scale=0.55]{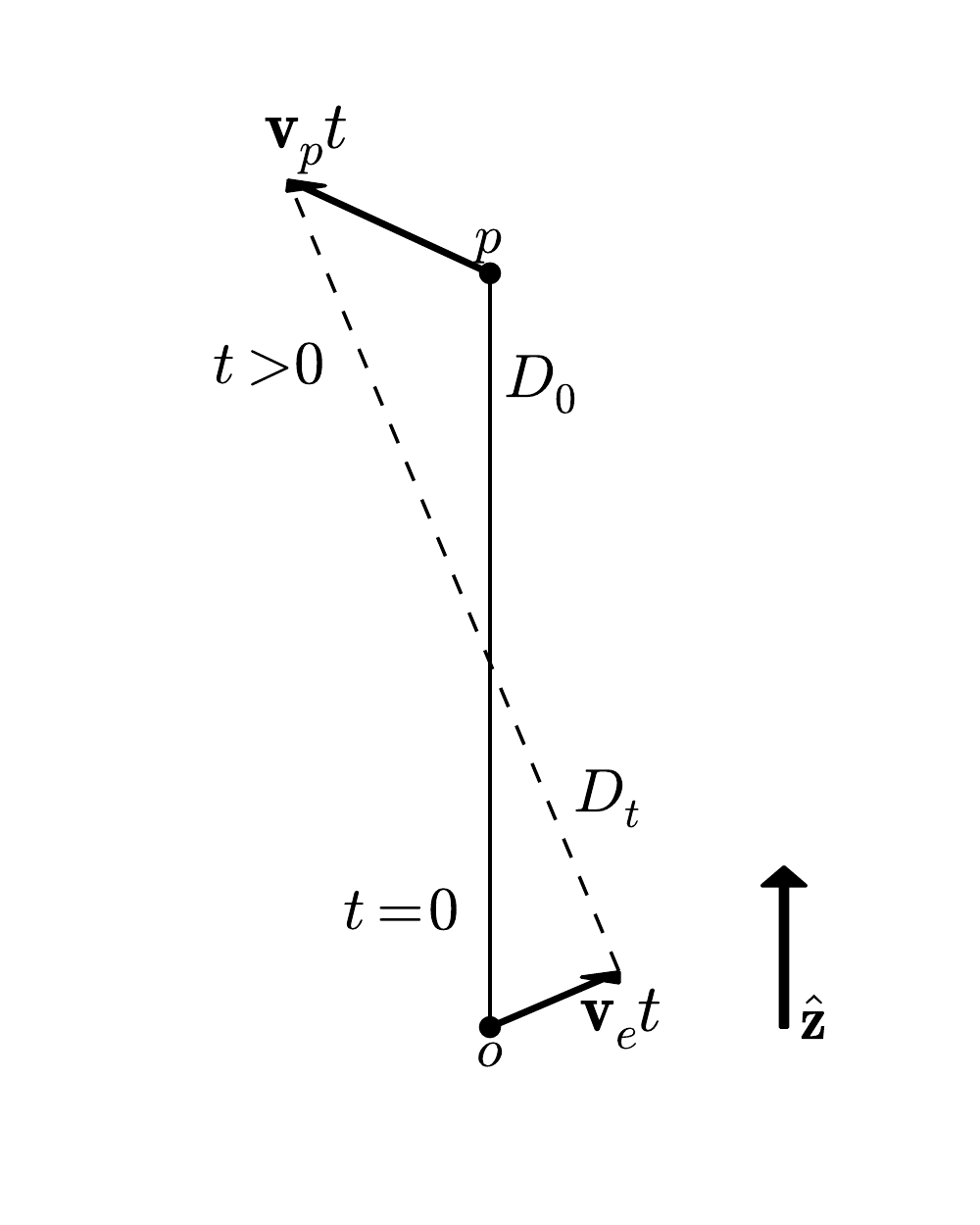} \hspace{0.5in}
\caption{\footnotesize 
Geometry showing change in LOS due to motion of pulsar $p$ and observer $o$.
DM is calculated by integrating along the $z$-axis taking into account the change in LOS. 
\label{fig:motions}
}
\vspace{5ex}
\end{figure}

In the following, we will develop the mathematical framework for variations in DM that we will use in following sections. Consider  changes in \DM\ that result from the relative motion of  the pulsar and observer, which  changes both the distance to the pulsar and  the direction of the LOS,
as shown in Figure~\ref{fig:motions}. For an initial pulsar position $\xpsrvec_0$ and Earth position $\xevec_0$, the initial distance $D_0 = \vert \xpsrvec_0 - \xevec_0 \vert$ increases (to first order in time) as
\be
D(t) \approx  D_0 + (\vpsrvec - \vevec)\cdot \nhat_0\  t \equiv D_0 + \Delta v_{\parallel} t,
\ee
where $\vpsrvec$ and $\vevec$ are the pulsar and Earth velocity vectors, respectively, $\nhat_0 = \Delta\xvec_0 / D_0 = (\xpsrvec_0 - \xevec_0)/D_0$ is the unit vector to the pulsar at $t=0$, and $\Delta v_{\parallel}$ is the apparent velocity of the pulsar parallel to the LOS. The next, quadratic term, $(\Delta v_\perp t)^2/2D_0$, where $\Delta v_\perp$ is the apparent velocity of the pulsar perpendicular to the LOS, is a factor $ \Delta v_\perp t/D_0 \sim 10^{-6}$ times smaller than the  linear term for typical parameters of $\Delta v\sim$ 100~km~s$^{-1}$ \citep{fk2006}, time span $T \sim 10$ years, and $D\sim$ 1~kpc \citep{NE2001}, and therefore can be ignored in calculating the distance. Conversely, the change in direction is determined by the transverse velocity
\be
\nhat(t) = \nhat_0  + D_0^{-1} \Delta \vvec_{\perp}t.
\ee

Let the initial LOS at $t=0$ be the $z$-axis and integrate over locations $\xvec_0(z) = z\hat \zvec$ to get the initial DM,
\be
\DM_0 = \int_{\zei}^{\zpi} dz\, n_e(\xvec_0(z)).
\ee
For $t>0$ we integrate over  a new interval $[\zet, \zpt]$ where 
\be
\zet= \zei + \vepar\ t, \quad \zpt = \zpi + \vppar\ t.
\label{eq:zs}
\ee
The sampled locations are now $\xvec(z,t) = \rvec(z,t) + z \zhat$ where $\rvec(z,t)$ is transverse to $\zhat$,
\be
\rvec(z,t) &=& \veffperpvec(z) t
\\
\veffperpvec(z) &=& \veperp + (\dvperp) \left(\frac{z-\zet}{\zpt-\zet}\right).
\label{eq:veff1}
\ee
The locations $\zet$ and $\zpt$ are evaluated at time $t$ and it is assumed that there is no significant acceleration correction over times of interest (weeks to decades). The effective transverse velocity, $\veffperpvec$, is a weighted sum of the pulsar's and Earth's velocities.  It is consistent with that given in Eq.~3 of \citet{cr98}, which also includes a term $-V_m$ for the velocity of the medium, and also with Eq.~C15 of \citet{1994MNRAS.269.1035G}.

The simplest approach is to evaluate the electron density for the $t>0$ LOS in terms of its values
for  the initial LOS,
\be
n_e(\xvec(z,t)) &=& n_e(\xvec_0(z)) +  [n_e(\xvec(z,t))  - n_e(\xvec_0(z))] \nonumber \\
	&\equiv& n_e(\xvec_0(z)) +  \Delta n_e(\xvec(z,t)).
\label{eq:nexpand}
\ee
The DM integral  over $[\zet, \zpt]$ can be expanded into integrals over the three intervals
$[\zei, \zpt]$, $[\zei, \zet]$, and $[\zpi, \zpt]$ to get 
\be
\DM(t) &=& \int_{\zet}^{\zpt} dz\, n_e(\xvec(z,t)) \nonumber \\
&=&
\int_{\zei}^{\zpi} dz\, n_e(\xvec(z,t)) +
\int_{\zpi}^{\zpt} dz\, n_e(\xvec(z,t))\nonumber \\
& & -\int_{\zei}^{\zet} dz\, n_e(\xvec(z,t)).
\label{eq:DMt1}
\ee
For the first integral we expand the integrand using Eq.~\ref{eq:nexpand} to get
\be
\int_{\zei}^{\zpi}\!\!dz\,n_e(\xvec(z,t)) = \DM_0 \!+\!\int_{\zei}^{\zpi}\!\!dz\,\Delta n_e(\xvec(z,t)). 
\ee
This gives
\be
\DM(t) & =& \DM_0\!+\!\int_{\zei}^{\zpi}\!\!dz\, \Delta n_e(\xvec(z,t)) \nonumber \\
& & +\!\int_{\zpi}^{\zpt}\!\!dz\, n_e(\xvec(z,t)) \!-\! \int_{\zei}^{\zet}\!\!dz\, n_e(\xvec(z,t)).
\label{eq:DMt2}
\ee
$\DM_0$ is the DM measured at time $t=0$, the first integral is the change in DM over the initial LOS (density fluctuation term), the second integral is the change in DM due to the pulsar's motion through its local environment (pulsar term), and the third integral is the change in DM due to the Earth/Solar System's motion through its local environment (Earth term). 

The integrand $\Delta n_e(\xvec(z,t))$ of the density fluctuation term needs to be considered only if 
 electron density variations are significant on length scales of order the offset between the LOS at $t$ and
the initial LOS at $t=0$, i.e., $\vert \Delta \xvec(z,t)\vert  = \vert\xvec(z,t) - \xvec_0(z)\vert \ll D_0$. For example, this offset $\ell \sim 20~{\rm AU}~\veffperp_{100} t_{\rm yr}$ for a fiducial velocity of 100~km~s$^{-1}$ and a year-long
time span. 
All evidence from the last few decades of interstellar scintillation studies are consistent with  there being variations 
on these (multiples of AU) and smaller scales \citep{Coles+1987,ars1995,Rickett+2000}. However, the detailed spectrum of variations on AU scales is not well known and appears to differ between the LOSs to different pulsars \citep{Stinebring+2000}.

The pulsar and Earth terms in Eq.~\ref{eq:DMt2} are over small intervals
$\zpt-\zpi = \vppar t$ and $\zet-\zei = \vepar t$ so, to first order in these intervals, the two terms
give $\nebar(\xvecbar_{p}) \vppar t$  and $\nebar(\xvecbar_{e}) \vepar t$, where $\nebar(\xvecbar_p)$  
and $\nebar(\xvecbar_e)$ are averages over the respective intervals centered on
$\xvecbar_p = \xvec_{p_0} + (v_{p_\parallel} t/2) \nhat(t)$ and 
$\xvecbar_e = \xvec_{e_0} + (v_{e_\parallel} t/2) \nhat(t)$, respectively. 
Unless there are large variations over the intervals, these average locations can be taken as the initial
ones at $t=0$. 
 The   \DM\  variations from these two terms are a simple consequence of   the change in pulsar distance   due to parallel motion because, as noted earlier, the transverse velocities enter only to second order and so are negligible in these terms.   

We  assume  that true {\em temporal} changes in electron density are negligible.  This is often a good assumption because  turbulent ISM velocities (of order a few km/s) are typically much smaller than pulsar velocities \citep{fk2006,frs2011}.   For slow pulsars and fast plasma screens (e.g., shock fronts),  the ISM velocity needs to be included and adds a term $-\vvec_m(z)$ (with $m$ for medium) to the effective velocity defined in Eq.~\ref{eq:veff1}.  For a purely turbulent medium, the velocity is stochastic and would depend on wavenumber.  However, a moving screen is easily described with a translational velocity.

\section{Linear Trends in DM}
\label{sec:linear_trends}


For a perfectly uniform medium with density $n_e$  the difference $ \Delta n_e(\xvec(z,t))$ vanishes and the total DM (found by combining Eqs.~\ref{eq:zs} and \ref{eq:DMt2}) is
\be
\DM(t) = \DM_0 + n_e (\vppar-\vepar)t, 
\ee 
giving a time derivative
\be
\frac{d\DM}{dt} & = & n_e (\vppar-\vepar)  \nonumber\\
&\approx &10^{-5}~ v_{100} {n_{e_{0.1}}}~{\rm pc~ cm^{-3}~yr^{-1}},  
\label{eq:dDMdt}
\ee
where the approximate estimate uses a  fiducial  relative velocity of 100~km~s$^{-1}$ and an electron density of 0.1~\cmthree \citep{frs2011}.  Observed DM derivatives range from approximately the nominal value in Eq.~\ref{eq:dDMdt}
up to values as large as $0.01$~pc~\cmthree~yr$^{-1}$, indicating that if the changing distance is the primary contribution to the observed trend, that 
the product $v_{100} {n_{e_{0.1}}}$ is as large as 1000. 

A slightly different form results for a medium with changes in density only on large length scales $\gg \vert\vppar-\vepar\vert t$, 
\be
\frac{d\DM}{dt} = n_e(\xvec_{p_0}) \vppar - n_e(\xvec_{e_0}) \vepar,
\label{eq:slow}
\ee
which indicates that changes in DM are affected by the local electron density on both ends of the LOS. For similar electron densities at the two locations, we expect the pulsar term to dominate because pulsar velocities are typically much larger than the Earth's orbital motion and the Sun's peculiar motion through the Local Interstellar Cloud (LIC), the latter about 28~km~s$^{-1}$ \citep{fk2006,frs2011}.  There will be exceptions, of course, for pulsars with low velocities or with small parallel velocity components.  

The LIC is about 2.5~pc across and has an internal gas density of $\approx 0.1-0.2$~\cmthree at a temperature of 7000~K \citep{frs2011}.   Assuming a completely ionized, uniform medium, the total DM through the cloud is at most  $\DM_{\rm LIC} \approx 0.5$~pc~\cmthree and the maximum derivative is 
\be
\frac{d\DM_{\rm LIC}}{dt}\ \approx 5.7\times 10^{-6}~ {\rm pc~ \cmthree~yr^{-1}}. 
\ee

The Earth's orbital motion is not relevant for the calculation of DM variations due to parallel motions because the Earth resides inside the heliosphere. A simplified form of Eq.~\ref{eq:DMt2} is therefore
\be
\DM(t) &=& \DM_0
	+ \left[n_e(\xvec_{p_0}) \vppar - n_e(\xvec_{e_0}) \vepar\right]t \nonumber \\
&&	+ \int_{\zei}^{\zpi} dz\, \Delta n_e(\xvec(z,t)). 
\label{eq:dmt}
\ee
However, the Earth's motion will matter when we later consider the interplanetary medium. In addition, the Earth term raises the interesting possibility that DM variations are partially correlated between different LOSs with an angular dependence that depends on the local ISM and on the direction of the Sun's peculiar velocity.

While linear trends in $\DM(t)$ have been recognized for many years, it is not {\it a priori} obvious whether they should be associated with the explictly linear term or with the density fluctuation term, which may quantify gradients transverse to the LOS. Some pulsars will show $\DM(t)$ variations where parallel motion is more important than transverse motion, and vice versa.  The two kinds of variations may be distinguishable.  If gradients and transverse motion are dominant, there should also be epoch-dependent refraction and flux-density variations on the same timescales.   However, parallel-motion effects need not be accompanied by strong modulations of scintillation parameters and flux densities because the structure of the ISM along the LOS will remain the same. We note that $\DM(t)$ will vary with time from parallel motion alone regardless of whether the ISM is uniform or not; no gradients in electron density $n_e$ are needed.   The variations will be monotonically increasing or decreasing with time if there is no transverse motion (of the pulsar, solar system, or medium). However, DM variations from transverse motion alone require gradients in $n_e$ that have components transverse to the LOS, i.e., $\nabla n_e \cdot \boldsymbol{r}(z,t) \ne 0$. Any gradients in $n_e$ will generally be manifested from both parallel and transverse motion. DM variations from parallel motion do not depend on the pulsar distance but the transverse change in LOS depends on location along the LOS, therefore influencing the observable effect from a transverse gradient. When the Earth's orbital velocity is important, such as for an MSP with low translational velocity, the contribution to $\DM(t)$ depends on $\nabla n_e \cdot \veffperpvec(z,t)$ and therefore will show a sinusoidal variation.

\section{Stochastic Variations in DM}
\label{sec:sf_and_spectrum}

\newcommand{\nerms}{{n_e}_{_{\rm rms}}}

Electron-density variations in the ISM can cause fluctuations in $\DM(t)$ that combine with the DM variations previously discussed. Many $\DM(t)$ time series have been shown to be consistent with purely stochastic variations in electron density; a list of references for epoch-dependent DM variations can be found in \citet{Lam+2015}. Following their treatment, we can describe the stochastic variations by a power-law wavenumber spectrum
\be
\Pne(\qvec) = \cnsq q^{-\beta}, \quad\quad q_1 \le q \le q_2,
\label{eq:pne}
\ee
where the wavenumber cutoffs, related to the inner and outer physical scales $\ell_2$ and $\ell_1$, respectively, are $q_1 = 2\pi/\ell_1$ and $q_2 = 2\pi/\ell_2$ and $\cnsq$ is the spectral coefficient. Eq.~\ref{eq:pne} assumes that the scattering irregularities are isotropic and the spectrum depends only on the magnitude of the wavenumber.  Evidence for anisotropic scattering exists along certain LOSs \citep[e.g.,][]{Brisken+2010} but the analysis is accordingly more tedious. The rms electron density is dominated by the largest scales ($\sim 1-100$~pc except in dense, compact regions) for $\beta > 3$ and $q_1 \ll q_2$. For a Kolmogorov medium, $\beta = 11/3$ \citep{r90}.

One useful statistic for quantifying DM variations is the DM structure function (SF),
\be
D_{\rm \DM}(\tau) & \equiv & \left\langle \left[ \DM(t+\tau) - \DM(t)\right]^2\right\rangle\nonumber\\
& = & \left\langle \left|\Delta^{(1)}\DM(t,\tau)\right|^2 \right\rangle,
\ee
where $\Delta^{(1)}\DM(t,\tau)$ is the first-order DM increment, because it removes any constant term and is closely related to the spectral index of the wavenumber spectrum when $\beta$ is in the scintillation regime (for wavenumbers $q_1 \ll q \ll q_2$ and $2 < \beta < 4$; \citealt{Lam+2015}). We can relate it to similar SFs found in the literature for  the electromagnetic phase perturbation imposed by the interstellar plasma $\phi$ and for the resulting dispersive time delay, $t$.  These are respectively $\phi = -c r_e \nu^{-1}\DM$, where $r_e$ is the classical electron radius, and  $t = d\phi/2\pi d\nu= K\nu^{-2}\DM$ with $K \equiv c r_e /2\pi$. We thus have
\be
D_t(\tau) =  K^2\nu^{-4} D_{\rm \DM}(\tau) = \left( 2\pi\nu \right)^{-2} D_{\phi}(\tau).
\label{eq:sfrelations}
\ee
The DM SF includes the effects of the systematic DM term due to the change in distance as well as the term involving the integrated difference $\Delta n_e(\xvec(t))$. Small scale, discrete structures on AU scales can contribute to  $\Delta n_e(\xvec(t))$  along with stochastic variations.

Together, discrete structures and the changing distance will produce contributions to the SF that are {\em quadratic} in $\tau$ and will contaminate the SF of the stochastic variations. A general feature of SFs is that they are quadratic when the lag $\tau$ is smaller than any characteristic timescale in the time series.   So for structures in the ISM with scale sizes $\ell$ of tens of AU that have characteristic crossing times $\ell /\veff \sim$~many years, quadratic SFs will be seen for lags of a few years or less.  For the case where only the distance-change term is relevant, $\DM(t+\tau) - \DM(t) \propto \tau$, it is easy to show that the SF is 
\be
D_{\rm \DM}(\tau) = \left[n_e(\xvec_{p_0}) \vppar - n_e(\xvec_{e_0}) \vepar\right]^2\tau^2.
\ee
More generally, if \DM\ variations are dominated by a linear gradient $d\DM/dt$, the SF is 
\be
D^{\rm (lin)}_{\rm \DM}(\tau) = \left[ \frac{d\DM}{dt} \right]^2\tau^2.
\ee
The SF of purely periodic variations in DM of the form $\DM(t) = C\cos(\omega t+\varphi)$ can easily be calculated as
\be
D^{\rm (per)}_{\rm \DM}(\tau) = C^2\left[1-\cos(\omega \tau)\right].
\ee

While the DM SF is typically calculated with time lags of days to years, it can be related to the implied phase SF on the diffractive interstellar scintillation (DISS) timescale of minutes to hours.  To do so, we use Eq.~\ref{eq:sfrelations} along with the fact that the 
scintillation timescale $\Dtiss$ corresponds to $D_{\phi}(\Dtiss) \equiv 1$~rad$^2$.  The corresponding DM SF value  (using $\lambda = c/\nu$) is 
\be
D_{\DM}(\Dtiss) &=& \left(\nu / 2\pi K \right)^2 D_{\phi}(\Dtiss) = \left(\lambda r_e\right)^{-2} \nonumber\\
	&=& 1.47\times 10^{-15} \nuGHz^2~({\rm pc~cm^{-3}})^2
\label{eq:D_DM_tau_dtiss}
\ee
Similarly
\be
D_t(\Dtiss) = \left(2\pi\nu\right)^{-2}  = 0.0253 \nuGHz^{-2}~{\rm ns^2}.
\ee
The SF can be extrapolated to larger time lags, and for the stochastic, Kolmogorov medium where $\beta = 11/3$,
\be
D^{\rm (sto)}_{\DM}(\tau) &=& (\lambda r_e)^{-2} \left(\tau / \Dtiss\right)^{5/3}
\label{eq:D_DM_kolmogorov}
\\
D^{\rm (sto)}_{t}(\tau) &=& (2\pi \nu)^{-2}  \left(\tau / \Dtiss\right)^{5/3}.
\ee

In general, the total DM SF can be written as the sum of the contribution from the systematic term and from the extrapolated stochastic term,
\be
D^{\rm (tot)}_{\DM}(\tau) & = &  D^{\rm (sys)}_{\DM}(\tau)  + D^{\rm (sto)}_{\DM}(\tau).
\label{eq:totalSF}
\ee
However, the systematic term cannot be separated as it will contain cross-terms if two or more components (e.g., linear plus periodic) are added together. For cases where the systematic term is significant, the time series for DM could be de-trended before calculating the SF, though de-trending can remove variance due to the electron density wavenumber spectrum. Equivalent to other discussions in the literature,  when a power-law wavenumber spectrum dominates electron-density variations the SF is essentially the SF of the density fluctuation term given in Eq.~\ref{eq:DMt2}.

We can relate the the DM SF from the random component to the rms of the DM variations. For a power-law wavenumber spectrum, the DM SF is
\be
 D_\DM(\tau) = 
 	 f_{\beta} \int_{z_{e_0}}^{z_{p_0}} dz\, \cnsq(z) \left[\veffperp(z) \tau\right]^{\beta  - 2},
\label{eq:dmsf}
\ee
where \citep [][Eq.  B6]{cr98}
\be
f_\beta = \frac{8\pi^2}{(\beta-2) 2^{\beta-2}}
\frac{\Gamma(2-\beta/2)}{\Gamma(\beta/2)}.
\ee
The numerical factor is $f_{11/3} = 88.3$ for a Kolmogorov wavenumber spectrum. Using the effective velocity of Eq.~\ref{eq:veff1} evaluated for the case where it is dominated by the pulsar velocity
and assuming $\cnsq$ is constant along the LOS, 
the \DM\ SF yields an rms \DM\ on a timescale $\tau$ for a Kolmogorov medium
\be
\sigma_{\DM} (\tau) &=& \left [ \frac{1}{2} D_\DM(\tau) \right ]^{1/2} \nonumber\\
	&=&  \left(\frac{\sqrt{3}f_{11/3}}{4}\right) \SM^{1/2} \left(\vpperp \tau\right)^{5/6}
	\nonumber \\
	&=& 1.9\times10^{-4}~{\rm pc~cm^{-3}}\times\nonumber\\ 
& &	\left(  \frac{\SM}{10^{-4}~\!{\rm kpc~m^{-20/3}}}  \right)^{\!1/2}\!\!\!\!\left({\vpperp}_{100} \tau_{\rm yr}\right)^{5/6}\!\!,
\label{eq:sigma_DM}
\ee
where the scattering measure (SM) is the LOS integral \citep{cl1991}
\be
{\SM} = \int_0^D ds\;C_n^2(s).
\ee
When the effective velocity is instead dominated by the Earth's velocity, as can be the case for some slow moving MSPs, the same expression applies but with $\vpperp$ replaced by $\veperp$. If both velocities are important, the integral in Eq.~\ref{eq:dmsf} needs to be evaluated explicitly.

Following Eq.~\ref{eq:sigma_DM}, an estimate of the rms DM gradient is
\be
\hspace{-2ex}
\sigma_{\rm d\DM/dt} &\approx& \frac{\sigma_{\DM} (\tau)}{\tau} \nonumber\\
& =& 1.9\times10^{-4}~{\rm pc~cm^{-3}~yr^{-1}} \times \nonumber\\
& &	\left(  \frac{\SM}{10^{-4}~{\rm kpc~m^{-20/3}}}  \right)^{1/2}\!\!{\vpperp}^{5/6}_{100} \tau_{\rm yr}^{-1/6}.
\label{eq:sigma_dDMdt}
\ee
The rms can be evaluated by using scintillation measurements to evaluate the scattering measure \SM\ \citep{cl1991} and by using proper motion measurements with distance estimates (from parallaxes or from \DM\ and a Galactic electron-density model) to estimate the pulsar velocity. 

One approach for comparing measured DM gradients with those expected from a Kolmogorov medium with no change in distance is to calculate the signal-to-noise-like ratio
\be
R_{d\DM/dt} = \frac{\left|d\DM/dt\right|}{ \sigma_{\rm d\DM/dt} }.   
\ee
When the gradient exceeds the prediction for a Kolmogorov model by a large factor, one of two interpretations may apply.   First, the medium may not have a Kolmogorov spectrum that encompasses both the small length scales that cause scintillation and the large 1--100~AU scales associated with DM variations.  Alternately, the excess derivative amplitudes can be caused by the changing pulsar distance as described above.     Identifying which of these interpretations apply requires consideration of other factors.    Transverse motions of the pulsar that cause the LOS to sample different irregularities will yield DM derivatives that are correlated with the absolute DM value whereas parallel motions that change the distance will not. 

Another approach compares the rms of the DM time series before and after the removal of a linear trend. Letting $\sigma_{\rm tot}^2 = \sigma_{\rm sto}^2 + \sigma_{\rm lin}^2$ be the total variance of the time series, we can define the ratio of rms after the removal of a linear trend to the rms before the removal as
\be
R_{\rm rms} = \frac{\left(\sigma_{\rm tot}^2 - {\hat{\sigma}_{\rm lin}}^2\right)^{1/2}}{\sigma_{\rm tot}}
\ee
where $\hat{\sigma}_{\rm lin}^2$ is the estimated variance of the linear trend. This definition restricts $0 \le R_{\rm rms} \le 1$. Realizations of DM time series will appear more linear when the wavenumber spectral index $\beta$ is large and the removal of the best-fit line for the time series will absorb low-frequency power from the frequency spectrum of DM. Conversely, when $\beta$ is low, the time series will appear closer to a white noise process, and the removal of a best-fit line will not change the resultant time series greatly.

We can solve for how $\sigma_{\rm sto}$ and $\sigma_{\rm lin}$ scale with observing time span $T$. Let the stochastic DM variation be a power spectrum $S_{\DM}(f) = A f^{-\gamma}$, where $A$ is a spectral coefficient related to $\Dtiss$ and $\gamma = \beta - 1$ (see Appendix~\ref{appendix:sfs} for more details). The variance is then
\be
\sigma_{\rm sto}^2 = \int_{f_1}^{f_2} S_{\DM}(f) df = \int_{f_1}^{f_2} A f^{-\gamma} df,
\ee
where $f_1$ and $f_2$ are the low- and high-frequency cutoffs, respectively, related to the wavenumber cutoffs $q_1$ and $q_2$. In the scintillation regime with $1 < \gamma < 3$, assuming $f_1 \ll 1/T \ll f_2$, the integral can be approximated as 
\be
\sigma_{\rm sto}^2 \approx \frac{A}{\gamma-1} T^{\gamma-1},
\ee
which for the Kolmogorov case implies $\sigma_{\rm sto} \propto T^{5/6}$. The variance from a deterministic, linear trend is 
\be
\sigma_{\rm lin}^2 & = & \frac{1}{T} \int_{-T/2}^{T/2} \left( \frac{d\DM}{dt} t\right)^2 dt\nonumber\\
& = & \frac{1}{12} \left(\frac{d\DM}{dt}\right)^2 T^2.
\ee
Therefore $\sigma_{\rm lin} \propto T$ and if a deterministic, linear trend is present, $\sigma_{\rm lin}$ will increase over $\sigma_{\rm sto}$ and $R_{\rm rms}$ will increase for longer observing timespans.

In addition to single, linear trends in DM, we can test for discrete changes in underlying linear trends in DM, such as from an ionizing bow shock (see \S~\ref{sec:bow_shocks}), versus stochastic changes from the turbulent medium by calculating the second-order increments of $\DM(t)$, $\Delta^{(2)}\DM(t,\tau) = \DM(t-\tau) -2\DM(t) + \DM(t+\tau)$, which remove linear components and relate to the curvature of the time series. The increments at a given $\tau$ will have a Gaussian distribution and deviations from this distribution will be indicative of structure other than from a turbulent medium. We can determine the variance in the distribution of increments at a given $\tau$, $\sigma^2_{\Delta^{(2)}\DM}(\tau)$, from the second-order DM SF, which can be written as
\be
D_\DM^{(2)}(\tau) &=& \left<\left|\Delta^{(2)}\DM(t,\tau)\right|^2\right>
\label{eq:second_order_SF}
\ee
For a Kolmogorov wavenumber spectrum, the second-order SF is related to the first-order SF, as well as the variance in the second-order increments, by
\be
D_\DM^{(2)}(\tau) = \sigma^2_{\Delta^{(2)}\DM}(\tau) \approx 0.8252 D_\DM^{(1)}(\tau).
\label{eq:increment_variance}
\ee
The derivation is provided in Appendix~\ref{appendix:sfs}. We can use Eqs.~\ref{eq:D_DM_kolmogorov} and \ref{eq:increment_variance} to analytically estimate the rms of the second-order DM increments given $\Dtiss$, which we use to analyze the slope changes in $\DM(t)$ for PSR B1534+12 in \S\ref{sec:interpretation_phenomena}.

\section{Refraction Effects and Timing}
\label{sec:refraction_effects}

Refraction of a radio point source by a high-density region in the ISM has been known to cause irregularities in electron-density time series. See \citet{1998ApJ...496..253C} and references therein for the case of a Gaussian plasma lens, to be considered shortly, and \citet{Coles+2015} for recent evidence of scattering events in pulsar timing data. One of the timing delays associated with refraction scale as $\nu^{-2}$ and is therefore degenerate with the dispersion delay, causing changes in the estimated DM. Consider a single ionized cloud that has characteristic scales  $\apar$ parallel and $\aperp$ transverse to the LOS and with a column density $\DM_c = \Nec \sim \nec \apar$ through the cloud along the LOS. The maximum phase change due to the clump is $|\phi_c| \sim \lambda r_e \Nec$ and the dispersion delay is
\be
{\dtDM}_c = \frac{\phi_{_c}} { 2\pi\nu} = \frac{\lambda^2 r_e \Nec}{ 2\pi c}.
\ee
The phase gradient across the LOS is then 
$\vert \nabla_\perp \phi \vert \sim \lambda r_e \Nec / \aperp$ and the refraction angle is
\be
\theta_{r_c} =  \frac{\lambda \vert \nabla_\perp \phi \vert }{2\pi}  \sim \frac{\lambda^2 r_e \Nec}{2\pi \aperp}
	\sim \frac{c{\dtDM}_c}{\aperp}.
\label{eq:refraction_angle}
\ee
There are two time delays introduced by refraction into barycentric arrival times.   The first is associated with the translation of topocentric TOAs by the propagation delay from the geocenter to the solar system barycenter. The direction to the pulsar is a key part of the translation, and refraction will induce an error in the barycentered arrival times.  Chromatic refraction causes the angle of arrival to differ from an assumed direction, implying a delay  \citep{fc90} that varies sinusoidally with an annual period and
an amplitude
\be
\dtbary_c  \sim \frac{r_{\oplus} \theta_{r_c}}{c} 
		\sim  \left(\frac{r_{\oplus}}{\aperp}\right)  {\dtDM}_c 
		\sim  \frac{{\dtDM}_c}  {\aperp_{\rm AU}},
\label{eq:dtbary}
\ee
where $r_{\oplus} = 1$~AU.
The second delay is the geometric increase in propagation path that is roughly
\be
\hspace{-2ex}
\dtgeo_c \sim \frac{D\theta_{r_c}^2}{2c} 
         \sim \frac{c D ({\dtDM}_c)^2}{2\aperp^2}
         \sim \frac{c D}{2r_{\oplus}^2} \left(\frac{{\dtDM}_c}  {\aperp_{\rm AU}}\right)^2\!\!.
\label{eq:dtgeo}
\ee
For a single clump, Eqs.~\ref{eq:dtbary} and \ref{eq:dtgeo} indicate that 
barycentric delay $\dtbary_c$ and geometric delay $\dtgeo_c$ are linear and quadratic, respectively, in the dispersion delay, ${\dtDM}_c$ ($\propto \nu^{-2}$ and $\nu^{-4}$, respectively). The barycentric and geometric delays are comparable  for pulsars within about 1~kpc
because $\theta_r \sim 1$~mas and $D\theta_r \sim 1$~AU, though there are wide variations of these values. 

\begin{figure}[t!]
\begin{center}
\includegraphics[width=0.52\textwidth]{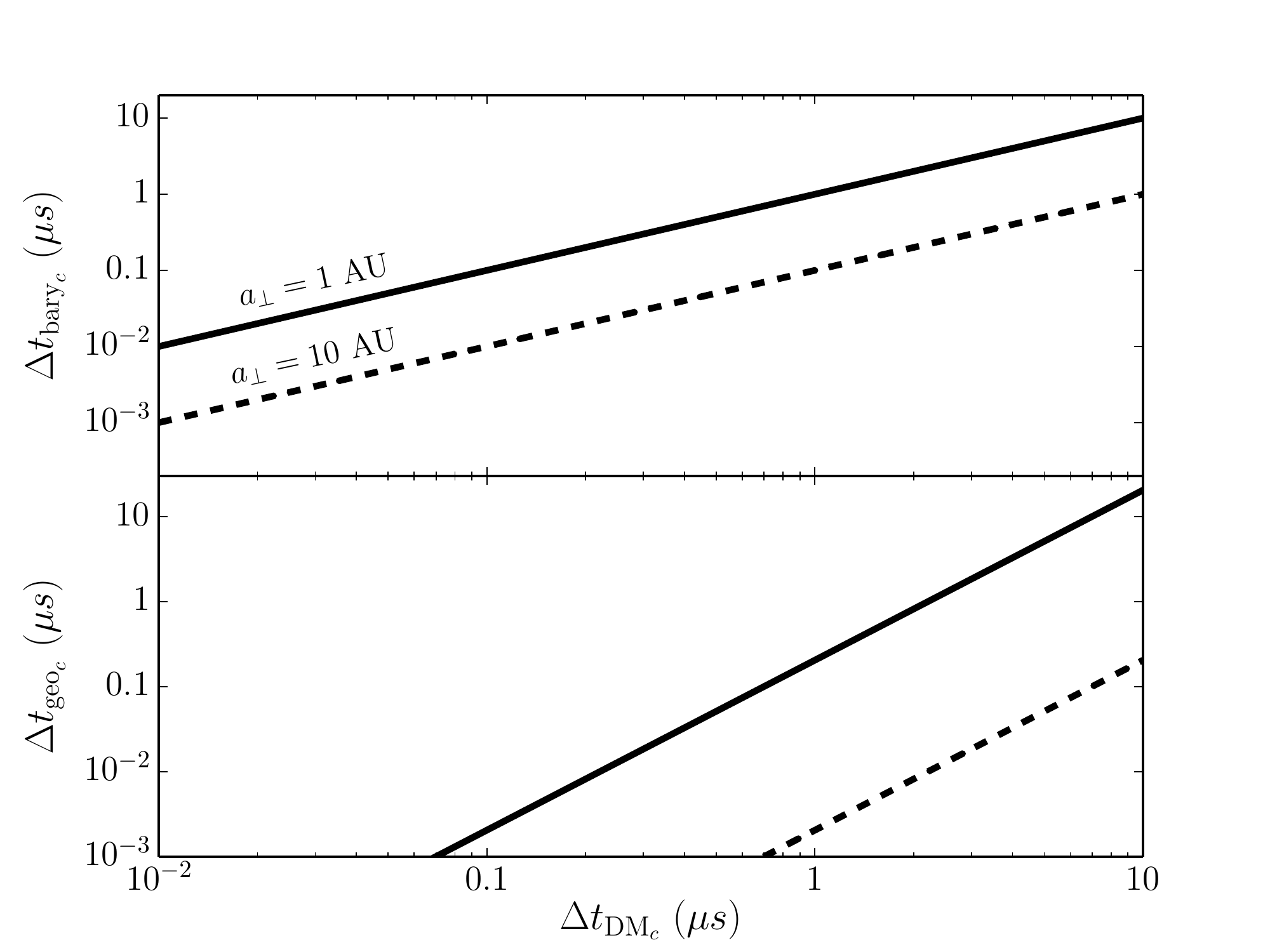}
\caption{\footnotesize Refraction delays plotted against DM delay from a single cloud at a distance $D = 1$~kpc. (Top) The barycentric delay for two clump scale sizes, as labeled. (Bottom) The geometric delay for the same two clump sizes. 
\label{fig:clump}
}
\end{center}
\end{figure}

Numerically, the refraction and dispersion delays are comparable for nominal parameter values
but any one of the three delays can dominate the the other two for reasonable distances and transverse
scale lengths, 
\be
\dtbary_c \sim 1~\mu s\, \left(\frac{{\dtDM}_{c, \mu s}}  {\aperp_{\rm AU}}\right),
\label{eq:dtbary_num}
\ee
and
\be
\dtgeo_c 	\sim 
		0.2~\mu s\, D_{\rm kpc} \left(\frac{{\dtDM}_{c, \mu s}}  {\aperp_{\rm AU}}\right)^2.
\label{eq:dtgeo_num}
\ee

Figure~\ref{fig:clump} shows $\dtbary_c$ and $\dtgeo_c$ plotted against ${\dtDM}_c$
for $D = 1$~kpc and for two transverse scale lengths ($\aperp = 1$ and 10~AU). 

A final consideration is multiple imaging.   \citet{1998ApJ...496..253C} analyze flux variations and caustics for an interstellar Gaussian plasma lens, i.e., a cloud with a Gaussian electron density profile.  The focal distance $D_{\rm f}$ of a clump is the minimum distance from the clump
at which  rays can cross,  
\be
D_{\rm f} \sim \frac{\aperp}{\theta_{r_c}} \sim \frac{\aperp^2}{c{\dtDM}_c} 
	\sim 2.4~{\rm kpc} ~ \aperp_{\rm AU}^2 {\dtDM}_{c, \mu s}.
\ee 
We therefore do not expect ray crossing and multiple images from nearby pulsars
 unless a clump is  small and dense.

\begin{figure*}[t!]
\epsscale{1.2}
\begin{center}
\plotone{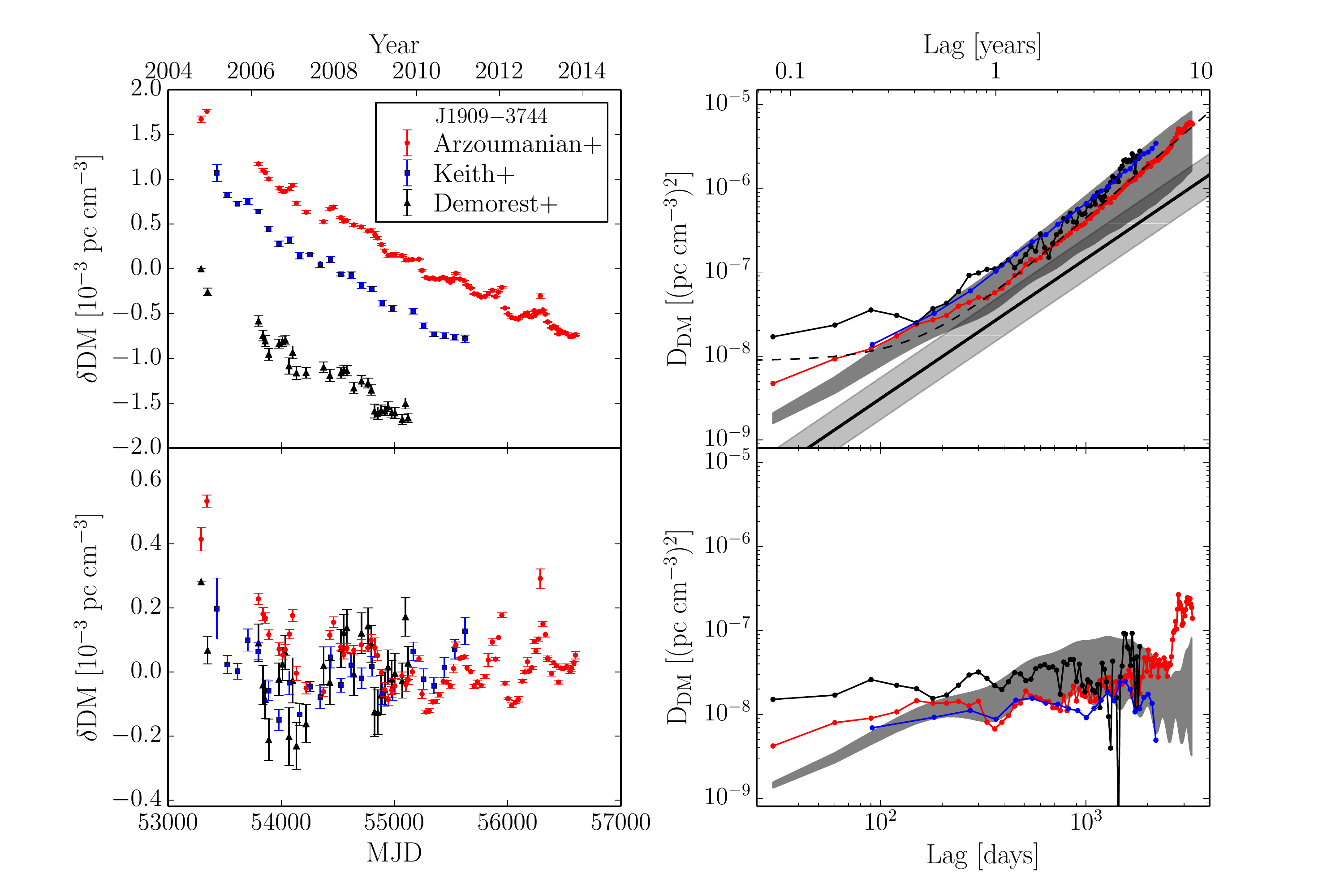}
\caption{\footnotesize Analysis of DM time series and SFs for PSR J1909$-$3744. Top left: DM offsets $\delta \DM(t)$ reported in \citet[red circles]{Arzoumanian+2015}, \citet[blue squares]{Keith+2013}, \citet[black triangles]{Demorest+2013}, see Figure~\ref{fig:smallplot} for more details. Top right: DM SFs for the three time series in the top left (with matching colors). The solid black line indicates the value of the SF inferred from the scintillation timescale and assuming DM variations only from a Kolmogorov wavenumber spectrum for the ISM. The light gray, hatched region shows nominal errors on the inferred SF from a multiplicative factor of $\sqrt{2}$ error on the scintillation timescale. The dark gray region indicates the $\pm 1 \sigma$ deviations from the mean SF for simulations of DM variations over nine years that include a Kolmogorov medium, a linear component from motion parallel to the LOS, a sinusoidal component, and measurement errors (see text for more information). Bottom left: DM offsets of the time series in the top left after a linear trend has been removed. Bottom right: DM SFs for the three time series in the bottom left. The gray region indicates the same as in the top right except that the best-fit linear trend has been removed from the simulated time series before calculating the SF.
\label{fig:J1909}
}
\end{center}
\end{figure*}

We can solve for the three time delays associated with refraction (dispersion, barycentric, geometric) by considering rays traveling through a Gaussian lens in the ISM. Following the treatment in \citet{1998ApJ...496..253C}, for a thin-screen approximation of the lens, the column density in two-dimensions can be written as
\be
N_{e_c}(\xvec) = N_0 \exp\left(-[|\xvec-\xvec_c|/a]^2\right),
\label{eq:cloud_density}
\ee
where $N_0$ is the maximum central column density and $a$ is the characteristic size of the lens. The screen phase $\phi$ is related to the electron density by
\be
\phi(\xvec) = -\lambda r_e \int_{\mathrm{screen}} dz\;n_e(\xvec,z) \equiv -\lambda r_e N_e(\xvec).
\label{eq:general_screen_phase}
\ee
For the Gaussian cloud, we therefore have
\be
\phi_c(\xvec) = -\lambda r_e N_0 \exp\left(-[|\xvec-\xvec_c|/a]^2\right).
\label{eq:general_screen_phase}
\ee
Using Eq.~\ref{eq:refraction_angle}, the refraction angle is 
\be
\bm{\theta}_{r_c}(\xvec) =  \frac{\lambda^2 r_e N_0}{\pi a^2} \xvec \exp\left(-[|\xvec-\xvec_c|/a]^2\right),
\ee
In general, the location of the incident ray paths on the Earth at location $\xvec_e$ intersecting the cloud at $\xvec_c$ from a pulsar at $\xvec_p$ must satisfy the equation
\be
\xvec_e = \xvec_c - \left[\bm{\theta}_{r_c}(\xvec_c) + \bm{\theta}_i(\xvec_p)\right]D 
\label{eq:ray_equation}
\ee
where $\bm{\theta}_i$ is the incidence angle of the pulsar rays  on the screen.

\section{Interpretation of Observed Pulsar Phenomena}
\label{sec:interpretation_phenomena}

\subsection{Linear Trends Versus Stochastic Variations}

\begin{figure*}[t!]
\epsscale{1.2}
\begin{center}
\plotone{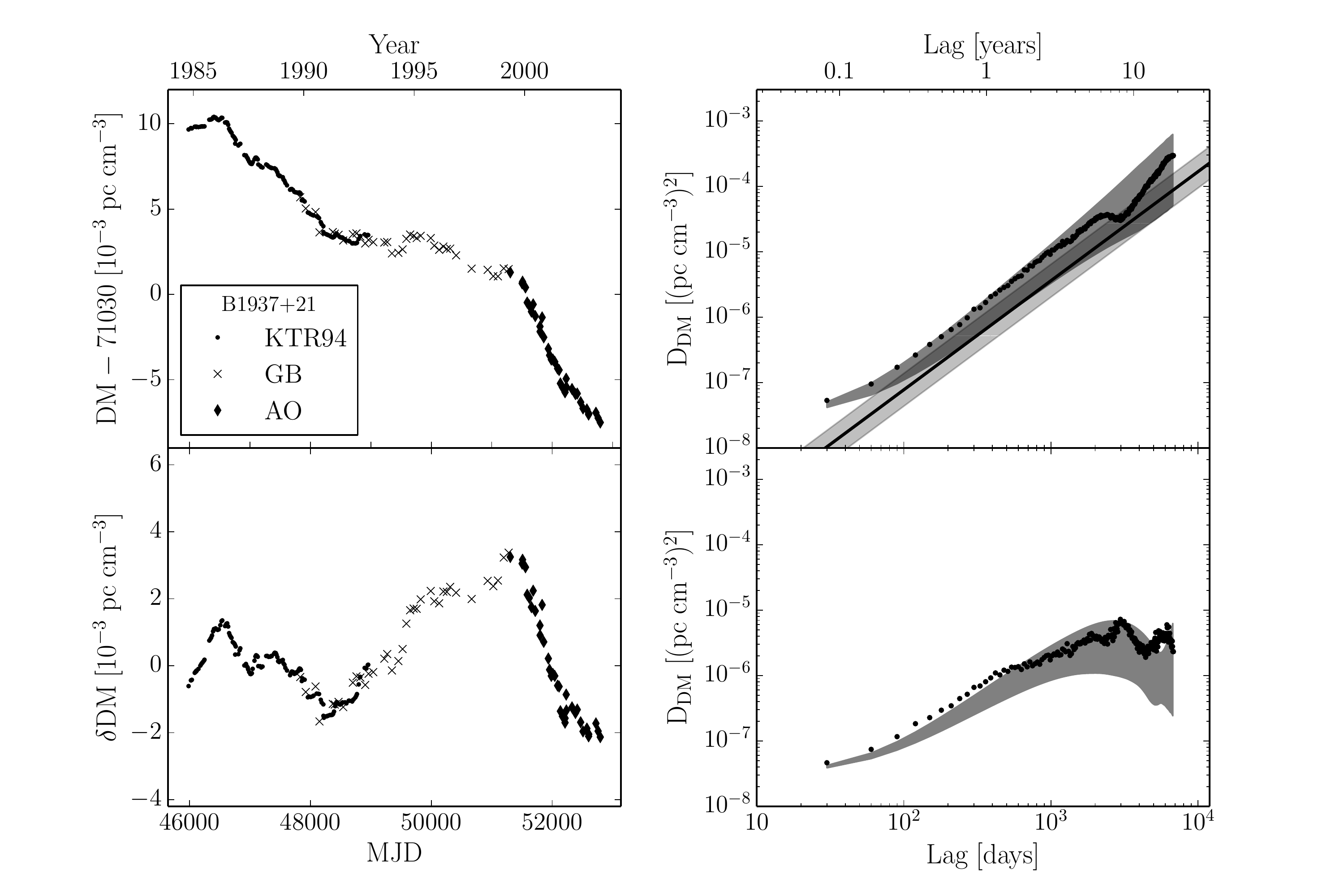}
\caption{\footnotesize Analysis of DM time series and SFs for PSR B1937+21. The format is the same as in Figure~\ref{fig:J1909}. The dots are from \citet{ktr1994} while the crosses and diamonds are from \citet{Ramachandran+2006} for the Green Bank (GB) 140-foot telescope and Arecibo Observatory (AO), respectively.
\label{fig:B1937}
}
\end{center}
\end{figure*}

\begin{figure*}[t!]
\epsscale{1.2}
\begin{center}
\plotone{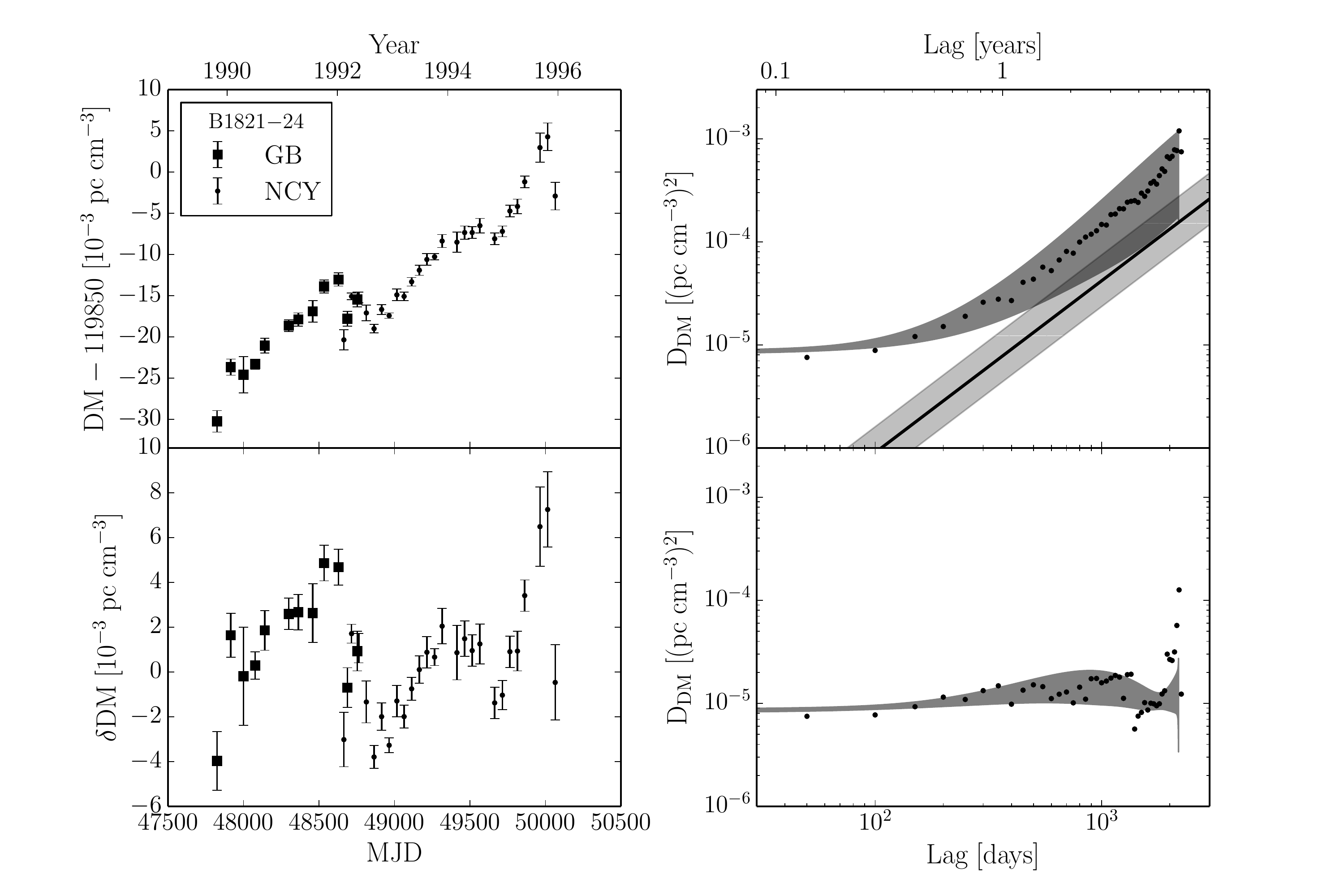}
\caption{\footnotesize Analysis of DM time series and SFs for PSR B1821$-$24. The format is the same as in Figure~\ref{fig:J1909}. The squares show measurements from the Green Bank (GB) 140-foot telescope \citep{b+93} while dots show measurements from the Nan\c{c}ay radio telescope (NCY) \citep{1997A&A...323..211C}.
\label{fig:B1821}
}
\end{center}
\end{figure*}

 We look at several examples of deterministic, linear DM trends seen in the literature below. To test our interpretations, we compare the time series against simulated DM variations with a Kolmogorov wavenumber spectral index following the same procedure as described in \citet{Lam+2015} by transforming scaled, complex white noise in the frequency domain to the time domain. The power spectrum of the electron density variations, $S_{\DM}(f) \propto f^{-\gamma}$, has a spectral index $\gamma = \beta - 1 = 8/3$ for the Kolmogorov case. The scalings of the coefficient of the power spectrum are consistent with the extrapolation of the SF $D_{\DM}$ by the scintillation timescale (see Eq.~\ref{eq:D_DM_kolmogorov}).

\subsubsection{PSR J1909$-$3744}

As described in \S2, the DM time series shows a decreasing, linear trend \citep{Demorest+2013,Keith+2013,Arzoumanian+2015} over a $\sim 9$-year timespan. \citet{Keith+2013} compare SFs of dispersion delay $D_t$ on long timescales with the extrapolations from the DISS timescale assuming a Kolmogorov spectrum. In several cases they find that the actual measurements exceed the extrapolation by large factors and that the slope of $D_t$ is larger than the Kolmogorov slope of $5/3$.  They conclude that the wavenumber spectrum is steeper than Kolmogorov.  In the case of J1909$-$3744, they find that the value of the DM SF for measured lags is about a factor of five higher than the extrapolation from the DISS timescale using a Kolmogorov scaling. 

We present an alternative interpretation that recognizes that the contribution to $\DM(t)$ from Kolmogorov fluctuations combined with transverse motion of the LOS can be contaminated by the changing distance between pulsar and Earth from parallel motion as discussed previously (see Eq.~\ref{eq:totalSF}). This contamination contributes a term to the SFs that scales as $\tau^2$, i.e., steeper than Kolmogorov, and that can dominate the overall amplitude of the SFs. 

The top left panel of Figure~\ref{fig:J1909} is identical to Figure~\ref{fig:smallplot} and shows the time series of the DM offsets, $\delta \DM(t)$, as reported in \citet{Demorest+2013}, \citet{Keith+2013}, and \citet[][NG9]{Arzoumanian+2015}. Again, the total measured DM is found by adding a constant $\DM_{\rm nominal}$ to the DM offsets, though the values will still differ due to other frequency-dependent parameters included in the timing models in each paper. Since the SF removes the mean, the differences are not important here. The bottom left panel shows $\delta \DM(t)$ after a linear trend has been removed. A periodic trend remains in the time series with a roughly one-year period. The panels on the right show the corresponding DM SFs of the time series on the left. In the top right, we show the extrapolation of $D_\DM$ assuming a purely Kolmogorov medium and using a scintillation timescale of 2258~s at 1.5~GHz \citep[solid black line,][]{Keith+2013} and note that all three SFs do lie well above this extrapolation. We calculate the SF for the \citet{Keith+2013} $\DM(t)$ with lag bins that are multiples of 0.25~yr (91.3~days), equal to the minimum sampling time for \citet{Keith+2013}. For the other two SFs, we use bins that are multiples of 30~days. The light gray, hatched region denotes nominal errors in the extrapolation from a multiplicative error of $\sqrt{2}$ on the scintillation timescale, known for PSR B1937+21 \citet{Cordes+1986,1990ApJ...349..245C,Keith+2013}. The dark gray region shows the $\pm 1 \sigma$ deviations from the mean DM SF on simulations of nine years of DM variations that include: a Kolmogorov wavenumber spectrum; the best-fit linear trend of the  data ($d\DM/dt = -2.27\pm 0.04 \times 10^{-4} \mathrm{\;pc\;cm^{-3}\;yr^{-1}}$); a sinusoid with a one-year period and an amplitude of $5 \times 10^{-5} \mathrm{\;pc\;cm^{-3}}$; and white, Gaussian noise with an rms of $\sigma_n = 2.4 \times 10^{-5} \mathrm{\;pc\;cm^{-3}}$ in the first five years and $1.2 \times 10^{-5} \mathrm{\;pc\;cm^{-3}}$ in the last four years, equal to the median error corresponding to each of the backends used in the NG9 data set. The DM SF from NG9 is consistent with these simulations. In the bottom right, we show the results of simulations when all of the above are included but a linear trend is fit and subsequently removed from the time series before computing the SF, which can remove power from both the linear component and some low-frequency structure in a given time series. The shape of the gray region matches some of the shape present in the SF, though the position indicates that the sinusoidal term should possibly have a smaller amplitude with a more peaked shape. Few numbers of DM increments in bins at large time lags lead to deviations from the mean SF. A more detailed analysis of the DM time series for J1909$-$3744 will be presented by M.\ L.\ Jones et al.\ (in preparation).

Using Eq.~\ref{eq:slow} and its assumption of density changes occurring at large length scales only, we can use our measured, best-fit $d\DM/dt$ to infer the electron density at the pulsar, $n_e(\xvec_{p_0})$. The transverse components of velocity are $v_\alpha = -50.61 \pm 0.01$~km~s$^{-1}$ and $v_\delta = -192.32 \pm 0.01$~km~s$^{-1}$ with a distance of $1.14^{+0.04}_{-0.03}$~kpc \citep{Jacoby2005,Antoniadis}. The barycentric, systemic radial velocity is $-37\pm 11$~km~s$^{-1}$ (J. Antoniadis, priv. comm.). We find the pulsar parallel velocity component $\vppar$ by removing the local solar motion and correcting for differential Galactic rotation. We take the local solar motion to be $18.0\pm0.9$~km~s$^{-1}$ in the direction $(l_g,b_g) = (47.9^\circ\pm3.0^\circ,23.8^\circ\pm2.0^\circ)$ and assume a locally flat, galactic rotation curve \citep{frs2011}. J1909$-$3744 lies nearly in the direction of the Galactic center with $(l_g,b_g) = (359.7^\circ,-19.6^\circ)$ and therefore the change in the velocity vector due to differential galactic rotation is negligible and we can ignore transverse components in our calculation. Taking the electron density of the LIC to be $n_e(\xvec_{e_0}) \approx 0.15\pm0.05$~\cmthree \citep{frs2011}, we find $n_e(\xvec_{p_0}) = 7.6\pm2.9$~\cmthree, about two orders of magnitude greater than the average local electron density of the galaxy in that region \citep{NE2001,frs2011}.

\subsubsection{PSR B1937+21}

\citet{2005AstL...31...30I} show a long-term trend in a 20-year DM time series extending to $\sim 2003.5$ (MJD 52800) that has a strong, decreasing trend with an average derivative $d\DM/dt \approx -1.14\pm 0.03 \times 10^{-3}$~pc~\cmthree~yr$^{-1}$. \citet{Ramachandran+2006} show similar results. Long-term correlated variations are superposed with the linear trend. The best-fit line of the SF is $\beta = 3.66 \pm 0.04$, though the analysis from \citet{ktr1994} on DM variations up to 1993 alone suggests $\beta = 3.874 \pm 0.011$. Both \citet{ktr1994} and \citet{Ramachandran+2006} fit the $D_\phi(\tau)$ to determine $\Dtiss$, which will be a biased estimator if a deterministic, linear trend is present.

We repeat our SF analysis as before, shown now in Figure~\ref{fig:B1937}, using the \citet{Ramachandran+2006} data. They include the time series from \citet[][circles]{ktr1994}, measured at 1400 and 2200~MHz. The crosses are measurements from the Green Bank (GB) 140-foot telescope between 800 and 1400~MHz, the diamonds are measurements from the Arecibo Observatory (AO) between 1400 and 2200~MHz. Differences in DM estimation and frequency-dependent delays used mean that we currently cannot align the \citet{Ramachandran+2006} time series with the \citet{Keith+2013} time series measured at later epochs. Therefore, we ignore the latter time series here and in subsequent analyses.

We measure $d\DM/dt = -8.39\pm 0.14 \times 10^{-4}$ pc \cmthree yr$^{-1}$ for the \citet{Ramachandran+2006} data, suggesting a long-term linear trend remains present in the time series. We again simulate a Kolmogorov medium with $\Dtiss = 327$~s at 1.5~GHz \citep{Keith+2013} and include a linear trend with slope measured above and additive, Gaussian white noise. The varying scintillation timescale over years is not included in the realizations and biases our overall results but not the conclusions. We find the rms of the noise by modeling the SF as $D_{\DM}(\tau) = C \tau^{\alpha} + 2\sigma_n^2$ (see Appendix~\ref{appendix:sf_error} for more details) and find $\sigma_n = 1.3 \times 10^{-4}$~pc~\cmthree. Again, the measured SF shows good agreement with our realizations.

\setcounter{footnote}{0}
\begin{center}
\begin{deluxetable*}{cccc|ccc|cc|cc}
\tablecolumns{11}
\tablecaption{Measurements of DM derivatives and $R_{d\DM/dt}$}
\tablehead{
\multicolumn{4}{c}{Pulsar Parameters} & \multicolumn{3}{c}{Scintillation Parameters$^{\rm a}$} & \multicolumn{2}{c}{DM Derivatives$^{\rm b}$} & \multicolumn{2}{c}{Derived Results}\\ \cline{1-11}
\colhead{Pulsar} & \colhead{DM$^{\rm c}$} & \colhead{PM$^{\rm c}$} & \colhead{$D^{\rm d}$} & \colhead{$\nu$} & \colhead{$\Dnuiss$} & \colhead{$\Dtiss$} & \colhead{$T$} & \colhead{$d\DM/dt^{\rm e}$} & \colhead{$\sigma_{d\DM/dt}$} & \colhead{$R_{d\DM/dt}$}\\
\colhead{} & \colhead{pc~\cmthree} & \colhead{mas~yr$^{-1}$} & \colhead{kpc} & \colhead{GHz} & \colhead{MHz} & \colhead{s} & \colhead{yr} & \multicolumn{2}{c}{10$^{-3}$~pc~\cmthree~yr$^{-1}$} & \colhead{}}
\startdata
J0358+5413 & 57.14 & $12.3\pm0.3$ & $1.1 \pm 0.2$ & 1.0 & 0.789 & - & 16.4 & --2.6 $\pm$ 0.8$^{\rm f}$ & 0.24 & 11.1\\
J0543+2329 & 77.71 & $22 \pm 8$ & 2.06 & 1.0 & 0.069 & - & 20.7 & --4.9 $\pm$ 0.6$^{\rm f}$ & 1.3 & 3.7\\
J0835$-$4510 & 67.99 & $57.98\pm0.08$ & $0.29 \pm 0.02$ & 0.61 & $1.5\!\times\!10^{-4}$ & 3 & 5.7 & 5 $\pm$ 1$^{\rm g}$ & 9.4 & 0.53\\
J1024$-$0719 & 6.49 & $59.7\pm0.3$ & $0.53 \pm 0.22$ & 1.5 & 268 & 4180 & 15.1 & 0.22 $\pm$ 0.06 & 0.047 & 4.7\\
J1045$-$4509 & 58.17 & $8.0\pm0.2$ & $0.30 \pm 0.17$ & 1.5 & 0.094 & 119 & 17.0 & --3.66 $\pm$ 0.13 & 0.90 & 4.1\\
B1534+12$^{\rm h}$ & 11.62 & $25.328\pm0.012$ & 1.051 $\pm$ 0.005 & 0.43 & 1.1 & 660 & 3.3 & --0.316 $\pm$ 0.010 & 0.081 & 3.9\\
 & & & & & & & 5.0 & --0.043 $\pm$ 0.008 & 0.076 & 0.57\\
 & & & & & & & 4.7 & --0.294 $\pm$ 0.007 & 0.077 & 3.8\\
 & & & & & & & 2.3 & 1.01 $\pm$ 0.03 & 0.086 & 11.7\\
 & & & & & & & 2.3 & --0.01 $\pm$ 0.05 & 0.086 & 0.12\\
J1543+0929 & 35.24 & $8.13\pm0.07$ & $7.7 \pm 1.2$ & 1.0 & 0.299 & - & 21.4 & 26 $\pm$ 5$^{\rm f}$ & 0.54 & 48.5\\
J1600$-$3053 & 52.33 & $7.2\pm0.3$ & $5.0 \pm 3.8$ & 1.5 & 0.09 & 271 & 9.1 & --0.63 $\pm$ 0.3 & 0.50 & 1.3\\
J1643$-$1224 & 62.41 & $7.3\pm0.3$ & $0.45 \pm 0.08$ & 1.5 & 0.022 & 582 & 17.0 & --1.23 $\pm$ 0.05 & 0.24 & 5.2\\
J1730$-$2304 & 9.62 & $20.27\pm0.06$ & 0.53 & 1.5 & 12.4 & 1615 & 16.9 & 0.56 $\pm$ 0.05 & 0.10 & 5.5\\
J1732$-$5049 & 56.82 & $9.9\pm0.3$ & 1.41 & 1.5 & 5.4 & 1200 & 8.0 & --0.88 $\pm$ 0.12 & 0.15 & 6.0\\
J1744$-$1134 & 3.14 & $21.02\pm0.03$ & $0.42 \pm 0.02$ & 1.5 & 60 & 2070 & 16.1 & --0.132 $\pm$ 0.018 & 0.084 & 1.9\\
J1833$-$0827 & 411 & $34\pm6$ & 4.5 & 1.0 & $1.6\!\times\!10^{-4}$ & - & 5.7 & --130 $\pm$ 20$^{\rm g}$ & 40.6 & 3.2\\
J1909+1102 & 149.98 & $9\pm8$ & 4.8 & 1.0 & 0.012 & - & 15.1 & --15.8 $\pm$ 1.2$^{\rm f}$ & 1.9 & 8.3\\
J1909$-$3744 & 10.39 & $37.10\pm0.02$ & $1.27 \pm 0.03$ & 1.5 & 37 & 2258 & 8.2 & --0.297 $\pm$ 0.006 & 0.087 & 3.4\\
J1935+1616 & 158.52 & $16.13\pm0.15$ & 4.55 & 0.61 & 0.002 & 18 & 34.1 & 2.3 $\pm$ 0.3$^{\rm f}$ & 1.6 & 1.5\\
B1937+21 & 71.02 & $0.421\pm0.003$ & $7.7 \pm 3.8$ & 1.5 & 1.2 & 327 & 15.5 & --0.59 $\pm$ 0.03 & 0.39 & 1.5\\
J2129$-$5721 & 31.85 & $13.3\pm0.1$ & $0.53 \pm 0.25$ & 1.5 & 17.1 & 3060 & 15.4 & --0.16 $\pm$ 0.04 & 0.061 & 2.6
\enddata
\footnotetext{Parameters for $\nu = 0.43$~GHz measurements from \citet{Bogdanov+2002}, for $\nu = 0.61$~GHz measurements from \citet{Stinebring+2000}, for $\nu = 1.0$~GHz measurements from PSRCAT \citep[][using $2\pi\Dnuiss\tauiss = C_1 = 1.16$]{PSRCAT}, and for $\nu = 1.5$~GHz measurements from \citet{Keith+2013}. The number of significant digits are provided by the individual references.}
\footnotetext{Timespan and $d\DM/dt$ references match.}
\footnotetext{Column data from PSRCAT \citep{PSRCAT} unless otherwise marked.}
\footnotetext{Distances with errors from parallax measurements (\url{http://www.astro.cornell.edu/research/parallax/} and references therein), distances without errors from NE2001 (errors are $\sim 20\%$), and distance for PSR B1534+12 from binary orbital period derivative \citep{fst2014}. }
\footnotetext{Values from \citet{Reardon+2016} unless otherwise marked.}
\footnotetext{\citet{Hobbs+2004}.}
\footnotetext{\citet{2013MNRAS.435.1610P}.}
\footnotetext{All values from \citet{fst2014} except the scintillation parameters \citep{Bogdanov+2002}.}
\label{table2}
\end{deluxetable*}
\end{center}

\subsubsection{PSR B1821$-$24}

\citet[][see also \citealt{b+93}]{1997A&A...323..211C} show a DM time series with a long-term increasing trend with $d\DM/dt \approx 0.005$~pc~\cmthree~yr$^{-1}$ over a six-year period. Again we ignore DM variations from \citet{Keith+2013} because of the absolute DM difference. Using measurements from GB and the Nan\c{c}ay radio telescope, \citet{1997A&A...323..211C} find that the spectral index of the wavenumber spectrum is $\beta = 3.727 \pm 0.211$. Figure~\ref{fig:B1821} shows our SF analysis, with red noise realizations with $\Dtiss = 75$~s at 1.5~GHz \citep{Keith+2013} and an estimated $\sigma_n = 2.1 \times 10^{-3}$~pc~\cmthree. The LOS to PSR B1821$-$24 is also consistent with a Kolmogorov medium.

\subsubsection{Deterministic Linear Trends from DM Derivatives}

We calculate $R_{d\DM/dt}$ for pulsars in the literature with measured $d\DM/dt$. To calculate $\sigma_{d\DM/dt}$, we use pulsars with a measured scintillation bandwidth $\Dnuiss$ or those that can be estimated from the scattering timescale $\tauiss$ using $2\pi\Dnuiss\tauiss = C_1$, where $C_1 = 1.16$ for a uniform medium with a Kolmogorov wavenumber spectrum. Assuming such a medium, we estimate the SM using Eq.~10 of \citet{NE2001} as
\be
\SM &=& 7.15\times10^{-4}~\mathrm{kpc\;m^{-20/3}} \nonumber\\
& & \times\left(\Delta \nu_{\rm ISS,MHz} \nuGHz^{-22/5} D_{\rm kpc}\right)^{-5/6}.
\ee
We either use parallax distances or binary orbital period derivative ($\dot{P}_b$) distances to estimate SM when available, and otherwise use DM distances from NE2001 \citep{NE2001}. We convert proper motion measurements into pulsar perpendicular velocities assuming $\vpperp$ dominates $\veffperpvec$ (which may not be true for slow moving MSPs) and differential galactic rotation is negligible, both of which may contribute systematic uncertainties in our analysis. However, under these assumptions, we combine SM, $\vpperp$, and the total observing span $T$ to calculate $\sigma_{d\DM/dt}$ and thus $R_{d\DM/dt}$. 

\begin{figure}[t!]
\hspace{-3ex}
\includegraphics[width=0.52\textwidth]{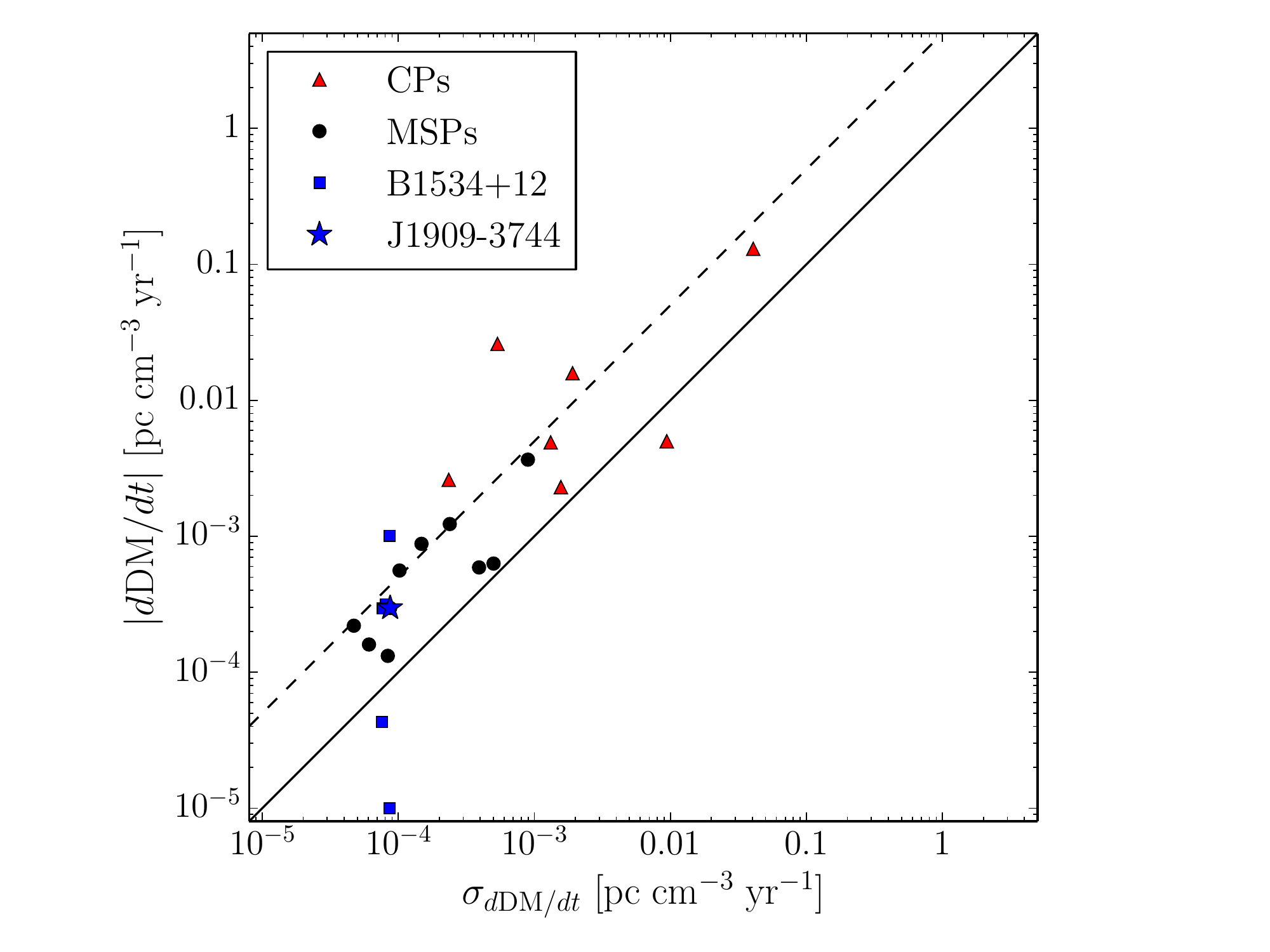}
\caption{DM time derivative $d\DM/dt$ versus the rms DM gradient $\sigma_{d\DM/dt}$. Canonical pulsars (CPs) are shown in red triangles and MSPs in black circles. We highlight the linear DM segments of the MSP B1534+12 in blue squares and the blue star for MSP J1909$-$3744. Two of the blue squares for B1534+12 closely overlap with J1909$-$3744. The solid line represents $R_{d\DM/dt} = 1$ whereas the dashed line represents $R_{d\DM/dt} = 5$, exponentially increasing to the top left.
\label{fig:R_dDMdt}
}
\end{figure}

\begin{figure}
\hspace{-3ex}
\includegraphics[width=0.52\textwidth]{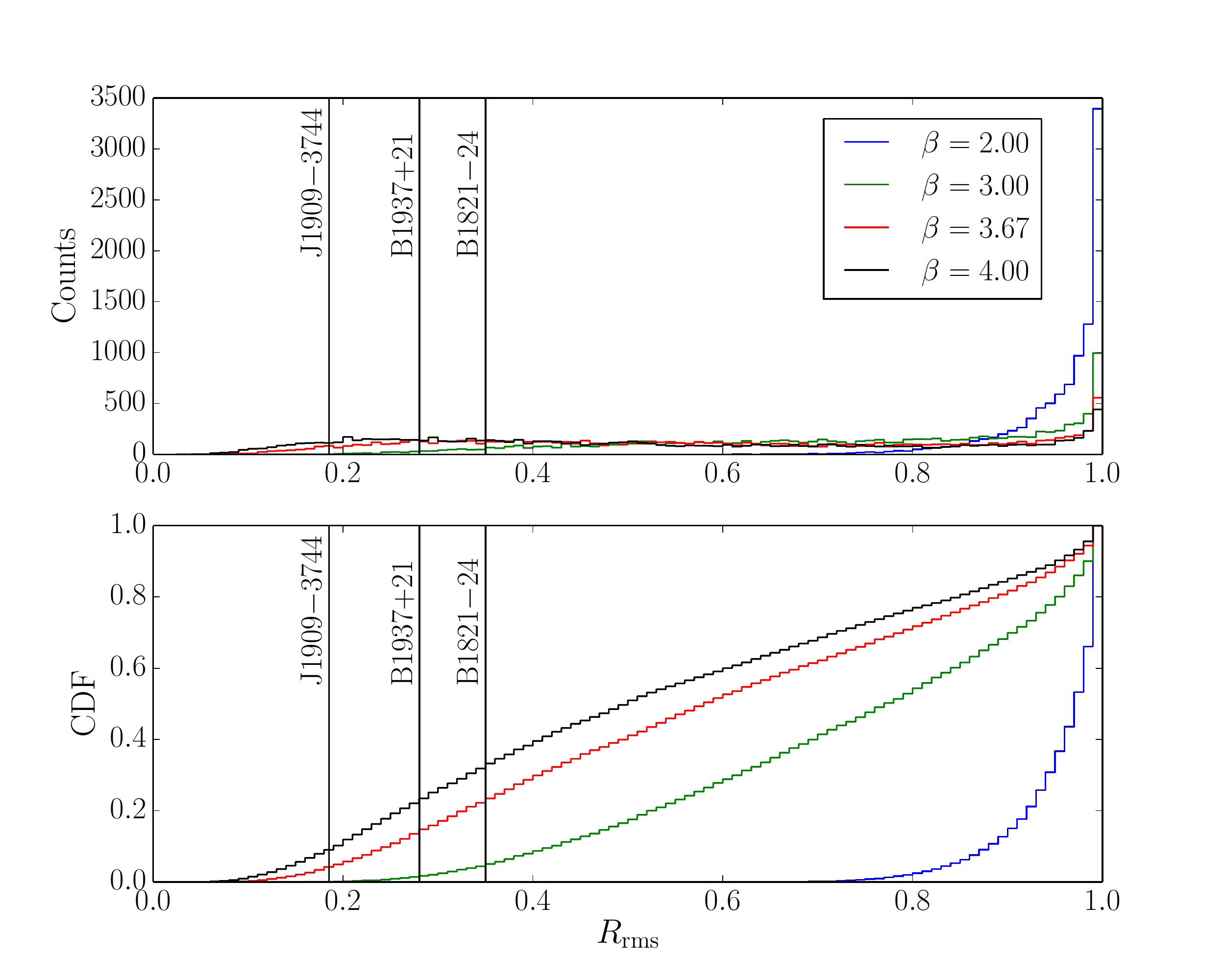}
\caption{Top: Histograms of the ratio $R_{\rm rms}$ for 10000 red noise process realizations with spectral index $\gamma = \beta - 1$. Bottom: Cumulative distribution functions of the histograms. The value of the $R_{\rm rms}$ is shown for the three pulsars in our analysis with single, linear trends. Lines towards the left indicate potential deviations from a wavenumber spectrum with spectral index $\beta$.
\label{fig:RMShistogram}
}
\end{figure}

Table~\ref{table2} lists pulsar values we use in the literature in our analysis and the results. Figure~\ref{fig:R_dDMdt} shows $|d\DM/dt|$ versus $\sigma_{d\DM/dt}$ for slow-period canonical pulsars (CPs, red triangles) and MSPs (black circles). We also highlight the five linear trends of the MSP B1534+12 (blue squares; discussed in the following section) and J1909$-$3744 (blue star). The lines represent $R_{d\DM/dt} = 1$ (solid) and 5 (dashed), where higher $R_{d\DM/dt}$ (increasing exponentially to the top left) is greater inconsistency with a Kolmogorov medium. J1909$-$3744 shows some evidence for deviations from a Kolmogorov medium. Several other pulsars show marginal or large deviations, including some MSPs with known chromatic timing noise such as PSR J1643$-$1224 in NG9 \citep[see also][]{NG9BWM}. Figure~\ref{fig:RMShistogram} shows the ratio $R_{\mathrm{rms}}$ (Eq.~34) for the DM time series of the pulsars examined so far, with J1909$-$3744 showing the most deviation (more toward the left) from a Kolmogorov spectral index, consistent with the conclusion from the $R_{d\mathrm{DM}/dt}$ metric.

\subsection{Non-Monotonic Trends from Electron-Density Structures in the ISM}

$\DM(t)$ time series from pulsars show a combination of linear trends, stochastic variations, and, in a few cases, fast changes in slope that are both positive and negative.    Apparent slope changes can appear in particular realizations of a stochastic process with a red power spectrum.  But they can also result from slab-like structures if they are suitably oriented relative to the LOS and the pulsar velocity.   Such slabs may represent static increases and deficits over the local mean electron density that contribute as the LOS changes with time.   Alternatively, they could be time-dependent owing to motions of the shock front through neutral gas.  Bow shocks produced by the pulsars  themselves may ionize atomic (and, less likely, molecular) structures as they move through the ISM.    

\citet{b+93} proposed that plasma wedges are responsible for linear trends in
$\DM(t)$.   A plasma wedge has linearly increasing column density $N_e(\xvec)$ transverse to the LOS. 
As the LOS moves across it, $\DM(t)$ will change linearly until the wedge boundary is reached, if
there is one.   The wedge will also refract by a constant refraction angle.    Unlike other structures,
however,  a wedge of this type will have zero transverse second derivative (except at the boundaries)
and therefore will not cause changes in measured flux density. 

\newcommand{\DMslab}{{\DM_{\rm slab}}}

\begin{figure}[t!]
\hspace{-7ex}
\includegraphics[width=0.60\textwidth]{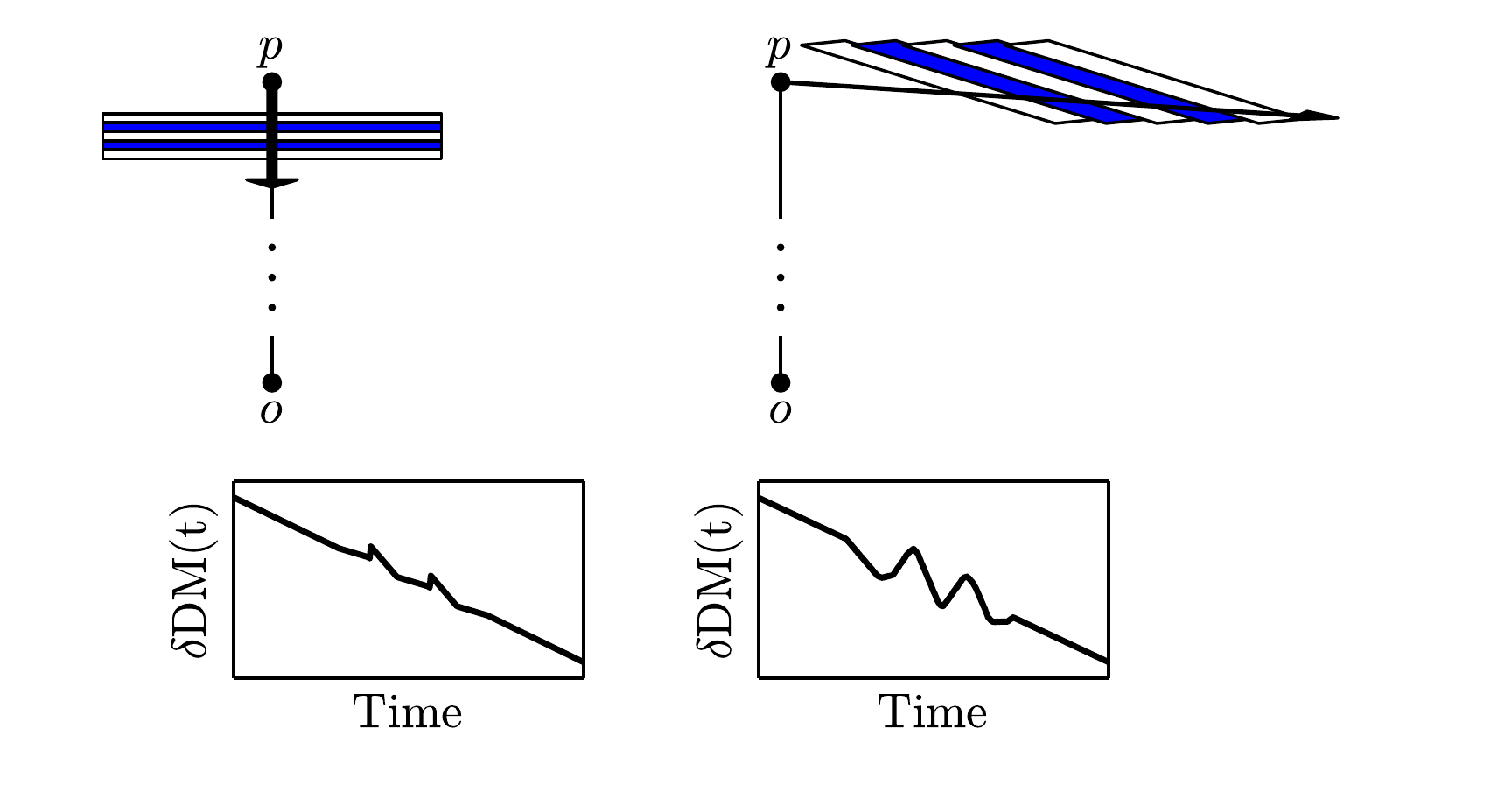}
\caption{Two cartoon geometries for a pulsar $p$ moving in different directions with respect to the line-of-sight between the pulsar at $p$ and the observer at $o$. Blue reprents high-density structures and white represents low-density structures. The bottom graphics show DM variations, $\delta \DM(t)$, that are monotonic on long timescales.
\label{fig:slabs}
}
\end{figure}

The effects of different geometries include:
\begin{quote}
{\em Transverse motion ($v_{p_{\parallel}} = 0$):}  For a density enhancement that is aligned with the LOS, $\DMslab(t)$ will consist of a positive-going `pulse' with duration equal to the pulsar travel time across the slab thickness.    For a density deficit (e.g., from encountering a slab of atomic gas), the pulse will be negative going. To first order, the pulsar distance does not change so the unperturbed DM is constant in time.

{\em Pulsar velocity component along the LOS ($v_{p_{\parallel}} \ne 0$) and aligned slabs:}  When the density slabs are aligned with the LOS,  $\DM(t)$ again show square-wave type pulses. The prevailing trend is for $\DM(t)$ to decrease as the pulsar distance gets smaller, but this is interrupted by the density deficits and enhancements.

{\em Pulsar velocity component along the LOS ($v_{p_{\parallel}} \ne 0$) and slanted slabs:}  
When the density slabs are 
slanted from the LOS,  $\DM(t)$ can show a saw-tooth pattern where it has a larger slope than the prevailing trend or a slope  with opposite sign. As in the previous case,  the prevailing trend is for $\DM(t)$ to decrease as the pulsar  distance gets smaller, but this is interrupted by the density deficits and enhancements.

{\em Pulsar velocity toward the observer ($v_{p_{\perp}}= 0$):}  
In this case, the pulsar can ionize atomic hydrogen as it passes into the slab. The \DM\ can increase even if the pulsar moves toward the observer and the prevailing trend is for a declining \DM.

\end{quote}

Some examples of the geometries can be seen in Figure~\ref{fig:slabs}.

\subsubsection{Ionized Bow Shocks}
\label{sec:bow_shocks}

So far, we have assumed that ISM structures are static.  However, the pulsar can actively modify its local environment.    An extreme case 
 is where the  pulsar's motion  toward the observer takes it
through  {\em atomic} hydrogen (HI) structures on scales of tens of AU and larger, including filaments, with a typical column density of order $N_{\rm HI} \sim 10^{-20}\;\mathrm{cm}^{-2}$ \citep[e.g.][]{2007ASPC..365...22S, 2007ASPC..365...59G, 2007ASPC..365...65M}.
  As the pulsar nears a filament, it will ionize the atomic
gas through a combination of radiation from the neutron star/magnetosphere and shock heating.   
The standoff radius of the bow shock is given by the balance of ram pressure and the pulsar's relativistic wind,
\be
r_s = \left( \frac{\dot E}{4\pi\rho v_p^2 c}\right)^{1/2}
	 \approx 266~{\rm AU}\, \dot E_{33}^{1/2}  n_{\rm HI}^{-1/2} v_{p_{100}}^{-1}
\ee
for $\dot E = 10^{33} \dot E_{33}$~erg~s$^{-1}$, a pulsar velocity in units of 100~km~s$^{-1}$, and an effective hydrogen density $n_{\rm HI}$~\cmthree. 
For the measured ranges of pulsar velocities, energy-loss rates ($\dot E$), and ISM densities, the standoff radius of the bow
shock is tens of AU to $\sim 0.1$~pc.   Therefore, $\DM(t)$ can show temporary increases even though the prevailing trend would be a decrease because of the decreasing distance.

Bow shocks will cause changes in DM only if the pulsar moves through a changing gas density (see, for example, PSRs J2124$-$3358 and B2224+65; \citealt{gjs2002,cc2004}). For completely ionized gas there may be a weak effect from the shock-enhanced gas density. A much larger effect will occur from neutral gas that is shock ionized or pre-ionized by radiation from shocked gas. The mean free path for ionizing radiation with $h\nu$ = 13.6 eV for a cross section $\sigma_{\rm HI} = 6.3 \times 10^{-18}$~cm$^{2}$ is
\be
l_{\rm HI} = (\sigma_{\rm HI} n_{\rm HI})^{-1} \approx \frac{1.1 \times 10^4~\mathrm{AU}}{n_{\rm HI,\cmthree}},
\ee
much larger than both the nominal standoff radius and the distance traveled by a pulsar in one year. However, for anticipated gas densities and temperatures (e.g., a shock temperature $T_s \approx 3 m_p v_p^2 /k \approx 3.6 \times 10^6$~K;
$m_p$ = proton mass, $v_p$ = pulsar velocity, $k$ = Boltzmann constant), there are insufficient photons to ionize a region of this size. This is why H$\alpha$ bow shocks are seen around some pulsars \citep[e.g., PSRs B1957+20, B2224+65, and J0437$-$4715;][]{br2014} in thin shells of pre-shocked atomic hydrogen that define the bow-shock’s contours.
For velocities $\sim 100$~km~s$^{-1}$ and densities $n_{\rm HI} \approx 1$ to 10 \cmthree, the distance traveled by the pulsar over years is less than or comparable to the standoff radius. The DM increment associated with shock ionized atomic gas is roughly
\be
\hspace{-3ex}
\delta \DM_{\rm bow} &\approx& \eta_s r_s n_{\rm HI} \nonumber \\
& \approx & 1.3 \times 10^{-3}~\mathrm{pc}~\cmthree \eta_s v_{p_{100}}^{-1}(n_{\rm HI} {\dot E_{33}})^{1/2},
\ee
where $\eta_s \sim 4$ is a factor that takes into account the compression of interstellar gas and its distribution inside the termination shock \citep{Clegg+1988}. The nominal value of $\delta \DM_{\rm bow}$ is sufficiently large to be interesting. Given the phase structure of the ISM, we expect that most pulsars will not reside in atomic gas but perhaps 40\% will \citep{Draine}.

\subsubsection{Small-Scale Electron Density Variations}

Because the ionized ISM contains a wide range of length scales \citep{ars1995}, the density fluctuation term  
in Eq.~\ref{eq:DMt2} involving $\Delta n_e(\xvec(t))$ also needs to
be considered.     Its contribution to DM is
\be
\delta \DM(t) = \int_{\zei}^{\zpi} dz\, \Delta n_e(\xvec(z,t)).
\label{eq:dDM1}
\ee
Typical scales transverse to the LOS  are 
$\vert\rvec\vert \sim  v_{\perp} t \sim 20~{\rm AU}\, v_{\perp 100}  t_{\rm yr}$.  The relevant velocity 
$\veffperpvec(z)$  is  largest at the pulsar position (c.f. Eq.~\ref{eq:veff1}) for cases where the proper-motion velocity is larger than the Earth's velocity. Elsewhere along the LOS and for slowly moving MSPs, the transverse scale can be substantially smaller.

  There is evidence for individual
structures in the ISM on AU  scales based on refraction effects in pulsar dynamic spectra,  extreme scattering events, and intraday variable sources.     These are likely confined to a small fraction of the LOS and will produce
maximum contributions to DM of order  $10^{-5} n_e \ell_{10\AU}$~pc~\cmthree where $\ell_{10\rm AU}$  is the path length through the structure.   The timescale for changes depends on the density, size, and velocity of the structure so the derivative $d\DM/dt$ can be comparable to or much smaller or larger than the contribution from the changing distance analyzed in the previous subsection. 

\subsubsection{Implications for PSR B1534+12}

\begin{figure*}[t!]
\epsscale{1.2}
\begin{center}
\plotone{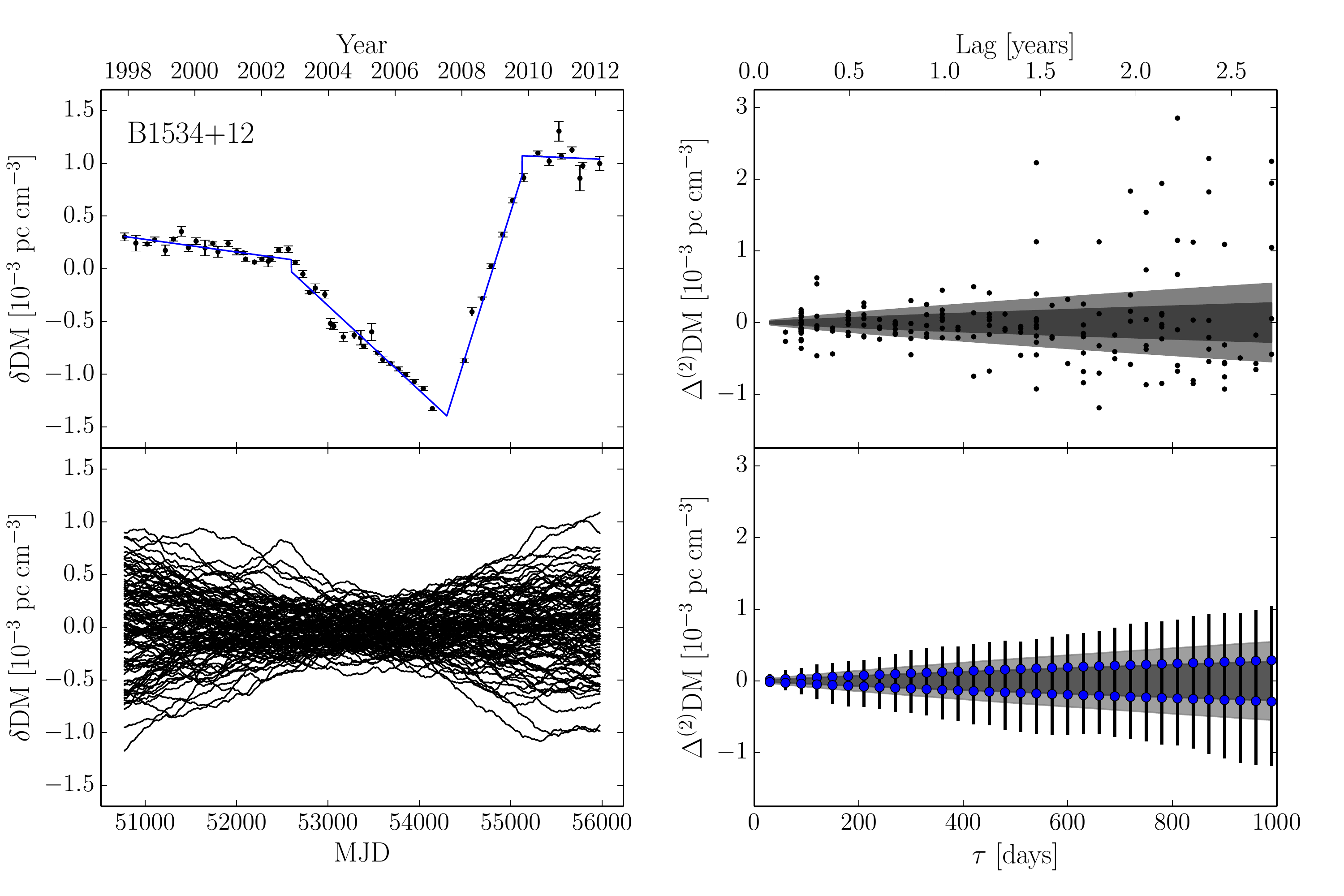}
\caption{\footnotesize Analysis of DM time series and SFs for PSR B1534+12. Top left: The DM offsets $\delta \DM(t)$ from \citet[][the first, isolated epoch has been removed]{fst2014} with their best fit linear trends overplotted. Top right: Second-order increments of DM, $\Delta^{(2)}\DM(t,\tau)$. The gray regions indicate the 1 and $2\sigma$ expected regions in Eq~\ref{eq:increment_variance} assuming a Kolmogorov wavenumber spectrum and the appropriate scintillation timescale, $\Dtiss = 660$~s. Bottom left: 100 realizations of $\delta \DM(t)$ from a Kolmogorov medium scaled to the appropriate scintillation timescale.  Bottom right: The second-order increments of the DM realizations in the bottom left. The shaded regions are the same as in the top right. The black bars indicate the range of increments as calculated from all of the realizations, with the blue circles indicating the $1\sigma$ bounds, matching the expectation.
\label{fig:B1534}
}
\end{center}
\end{figure*}

We make use the second-order SF approach developed in \S\ref{sec:sf_and_spectrum} to analyze the DM time series presented in \cite{fst2014}. While they note five significant linear trends in DM, they have no temporal information in the first block and so we remove it from our analysis. 

Figure~\ref{fig:B1534} shows the DM time series along with the second-order increments of DM, which we use to indicate the presence of discrete changes in linear trends over the expectation from a purely Kolmogorov medium. To account for the unequal sampling in $\delta \DM(t)$, we calculate increments as a function of $\tau$ by finding two points separated from the central time $t$, one within the range $\tau\pm\tau/2$ and the other within $-\tau\pm\tau/2$. Increasing $\tau$ by 30 days at a time, we plot the second-order increments in the top right panel. Combining Eqs~\ref{eq:D_DM_kolmogorov} and \ref{eq:increment_variance}, the probability distribution of the increments will be a Gaussian function. Using $\Dtiss = 660 \pm 180$~s measured at 430~MHz \citep{Bogdanov+2002}, the 1 and 2$\sigma$ expected regions are shown in the gray bands. The points in the top right of the plot outside of the bands and at $\tau > 500$~days all result from the concave up turnover between the second and third linear components and represent a $\sim 7-14\sigma$ deviation from the expectation of a purely Kolmogorov medium. Points well below the bands come from either of the other two changepoints. The points deviating from zero at low lags are purely from the noise in the measurements, not accounted for in the gray bands. The bottom left shows 100 realizations of DM purely from Kolmogorov power-law wavenumber spectra and the full range (minimum to maximum value) of second-order increments for each $\tau$ on the bottom right from all 10,000 realizations. We show the results of our simulations to demonstrate that there is good agreement between the rms of the second-order increments from simulations and the analytic solution. Again, the fact that several measured increments for PSR B1534+12, notably the ones associated with the second changepoints, fall outside of the expectations from simulations imply that the upturn in DM cannot be due to a purely Kolmogorov medium. The time series is similar to those shown in Figure~\ref{fig:slabs} after a linear trend has been removed, suggesting that there are interleaved density structures along the LOS.   Contemporaneous scintillation parameters and pulsar flux density measurements would be valuable for testing whether the \DM\ time series is at all contaminated by diffraction and refraction effects.

\section{Periodic Variations in DM}
\label{sec:annual_trends}

In this section, we determine how periodic trends can appear in DM time series. $\DM(t)$ will vary as the LOS passes across spatial gradients in electron density. Local electron-density variations in time will also cause DM variations. We assess the periodicites and phases associated with each periodic contribution to DM.

\subsection{Ionosphere}

\begin{figure*}[t!]
\epsscale{1.2}
\begin{center}
\plotone{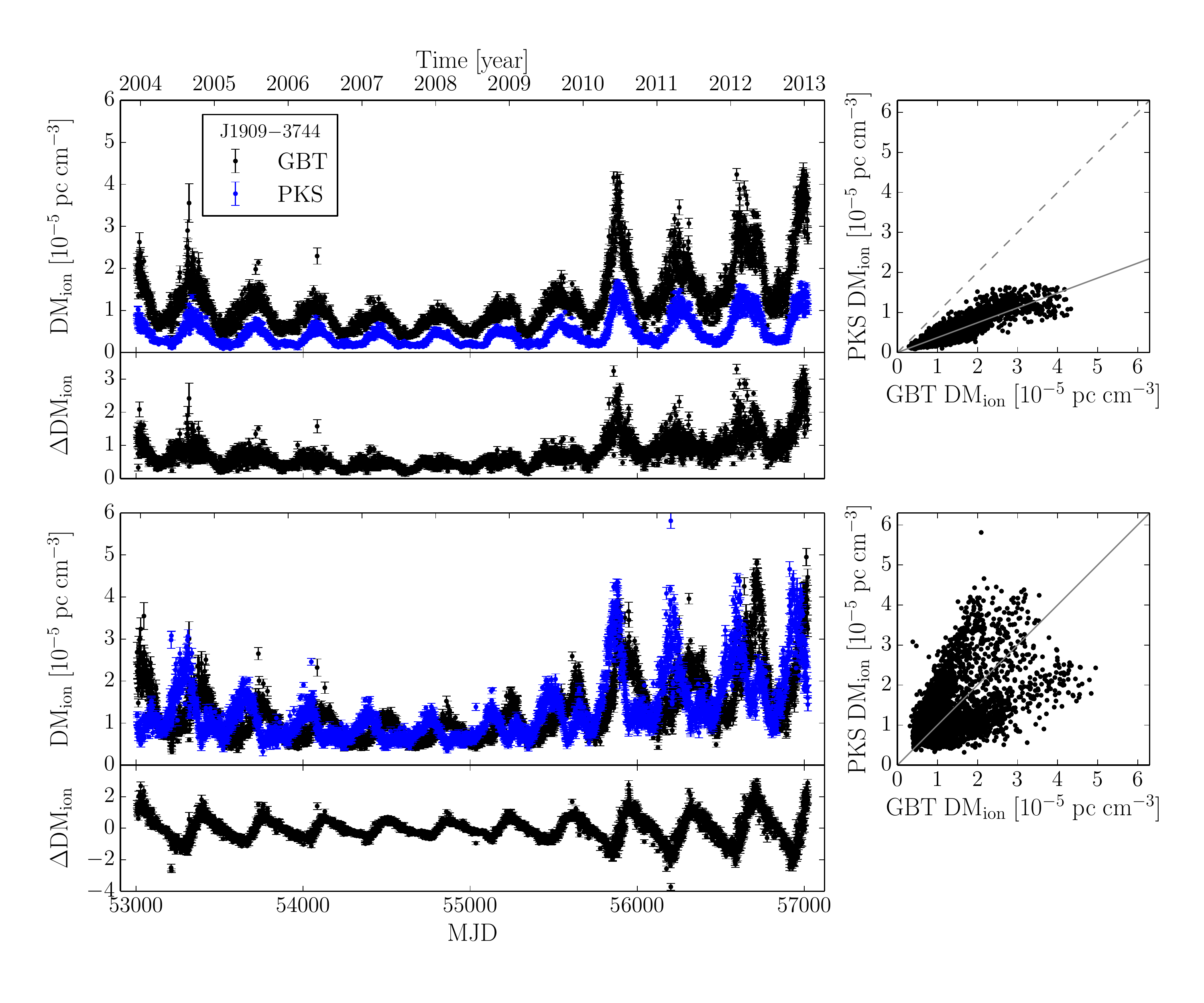}
\caption{Estimates of the ionospheric contribution to DM from interpolated global navigation system satellite measurements along the LOS to J1909$-$3744 from Green Bank Telescope (GBT, black) and the Parkes Telescope (PKS, blue) daily over nine years. Top left: Measurements of DM computed daily when J1909$-$3744 transits each telescope. The difference between the GBT and PKS DM estimates are shown in the panel beneath. Top right: DM estimated from Parkes at transit versus from GBT at transit. The dashed, diagonal line represent the same value at both sites while the solid line is the difference in the geometric factor $G(\theta_z)$ between both sites. Bottom: Similar to the top  except that DM estimates were determined simultaneously, approximately two hours after transit at GBT so that $G(\theta_Z)$ was the same at both sites. 
\label{fig:J1909ionosphere}
}
\end{center}
\end{figure*}

Changes in the electron-density within the ionosphere can cause differences in DM between observatories. The changes correlate with the incident solar flux at a particular location. Variations are known to occur daily from Earth's rotation, yearly due to Earth's orbital motion, and on 11-year cycles due to changes in solar magnetic activity \citep[see][for an ionospheric electron-density model over a specific LOS]{hr2006}. Measurements of the electron density can be peformed by satellite, rocket, incoherent scatter radar, and ionosonde.

The ionosphere can be represented as a series of semi-Epstein layers \citep{Rawer1982} with electron density as a function of the normalized distance parameter $z$,
\be
n_e(z) = \frac{4 n_0}{\left[1+\exp\left(z\right)\right]^2} \exp\left(z\right)
\label{eq:epstein_layer}
\ee
with the peak electron density of the layer, $n_0$, and $z = (h_s-h_0)/H$, where $h_s$ is the height above the Earth's surface, $h_0$ is the height of the peak electron density, and $H$ is the characteristic thickness of the layer. Note that as $h_s \gg h_0$, $n_e(z)$ tends towards zero.

 \citet[][NeQuick 2\footnote{\url{http://t-ict4d.ictp.it/nequick2}}]{ncr2008} model the E, F1, and F2 ionospheric layers using one semi-Epstein layer to describe the bottomside and topside of each layer. They introduce a ``fadeout'' function that multiplies $z$ in the E and F1 layer functions to prevent secondary maxima around the F2 peak height. The peak heights, peak electron densities, and thicknesses of each layer change as a function of latitude and longitude, $(\phi_g,\lambda_g)$, over the Earth's surface due to the structure of its time-varying magnetic field. There are additional time-dependent factors regarding the incident solar flux at a given $(\phi_g,\lambda_g)$, including the change in the Sun's zenith angle over a day, the change in the seasons for a given latitude $\phi_g$, and the variable solar flux that changes both daily and over a solar cycle. Therefore, the three layer parameters, $n_0$, $h_0$, and $H$, are all complex functions of longitude, latitude, and time.

In general, the DM is the integral of electron density over some path $s$ through the atmosphere that depends on the geographic coodinates of the observatory and the apparent coordinates (i.e., altitude and azimuth) of the pulsar, which in turn depend on the equatorial coordinates of the source $(\alpha_e,\delta_e)$ and time $t$. The path to integrate over is then $s(\phi_g,\lambda_g,\alpha_e,\delta_e,t)$ and the total ionospheric contribution to DM is simply the line integral 
\be
& & \DM_{\mathrm{ion,NeQuick}}(t,\phi_g,\lambda_g,\alpha_e,\delta_e)\nonumber  \\
& & = \int_0^{s_{\rm max}}\!\!\!\!\!\! \sum_{i=\mathrm{E,F1,F2}} \!\!\!\!\!\! n_{e,i}(s(t,\phi_g,\lambda_g,\alpha_e,\delta_e),\phi_g,\lambda_g,t) ds
\label{eq:DMionosphere1}
\ee
up to some maximum distance $s_{\rm max}$, where we sum the total electron over each layer.

We study the variations of electron density in the ionosphere using two methods. The first estimates the ionospheric contribution to DM using global navigation satellite system (GNSS) measurements from the International GNSS Service \citep[IGS;][]{IGS}. The total electron content (TEC) is measured via frequency-dependent signal propagation delays similarly to pulsar timing delays but between a ground receiver and a transmitting satellite along a given LOS. Using multiple LOSs at a given time, the IGS constructs a 2D surface map of the ionospheric electron density. These maps typically have time resolution of two hours and spatial resolution of 2.5 $\times$ 5.0 degrees in latitude and longitude, respectively. We linearly interpolate intermediate TEC values in both space and time. While the original measurements between receiver and satellite are along some altitude and azimuth, the reported TEC values are in the zenith direction. Therefore, we must adjust the measurements for a particular LOS. To simplify, we approximate the ionosphere as a uniform slab of electrons with an inner ($s_{\mathrm{min}}$) and outer ($s_{\mathrm{max}}$) height of 60 and 600 km above the Earth's surface, respectively. Therefore, we can estimate the TEC along a LOS by multiplying the zenith TEC by a geometric factor $G(\theta_z)$ that accounts for the increase in path length through the ionosphere and depends only on the zenith angle $\theta_z$ to the pulsar, 
\begin{align}
& \DM_{\mathrm{ion,IGS}}(t,\phi_g,\lambda_g,\alpha_e,\delta_e)\nonumber \\
&   = n_e (\phi_g,\lambda_g,t)(s_{\mathrm{max}}\!-\!s_{\mathrm{min}}) G(\theta_z(\phi_g,\lambda_g,\alpha_e,\delta_e,t)).
\label{eq:DMionosphere2}
\end{align}

\begin{figure}[t!]
\begin{center}
\includegraphics[width=0.52\textwidth]{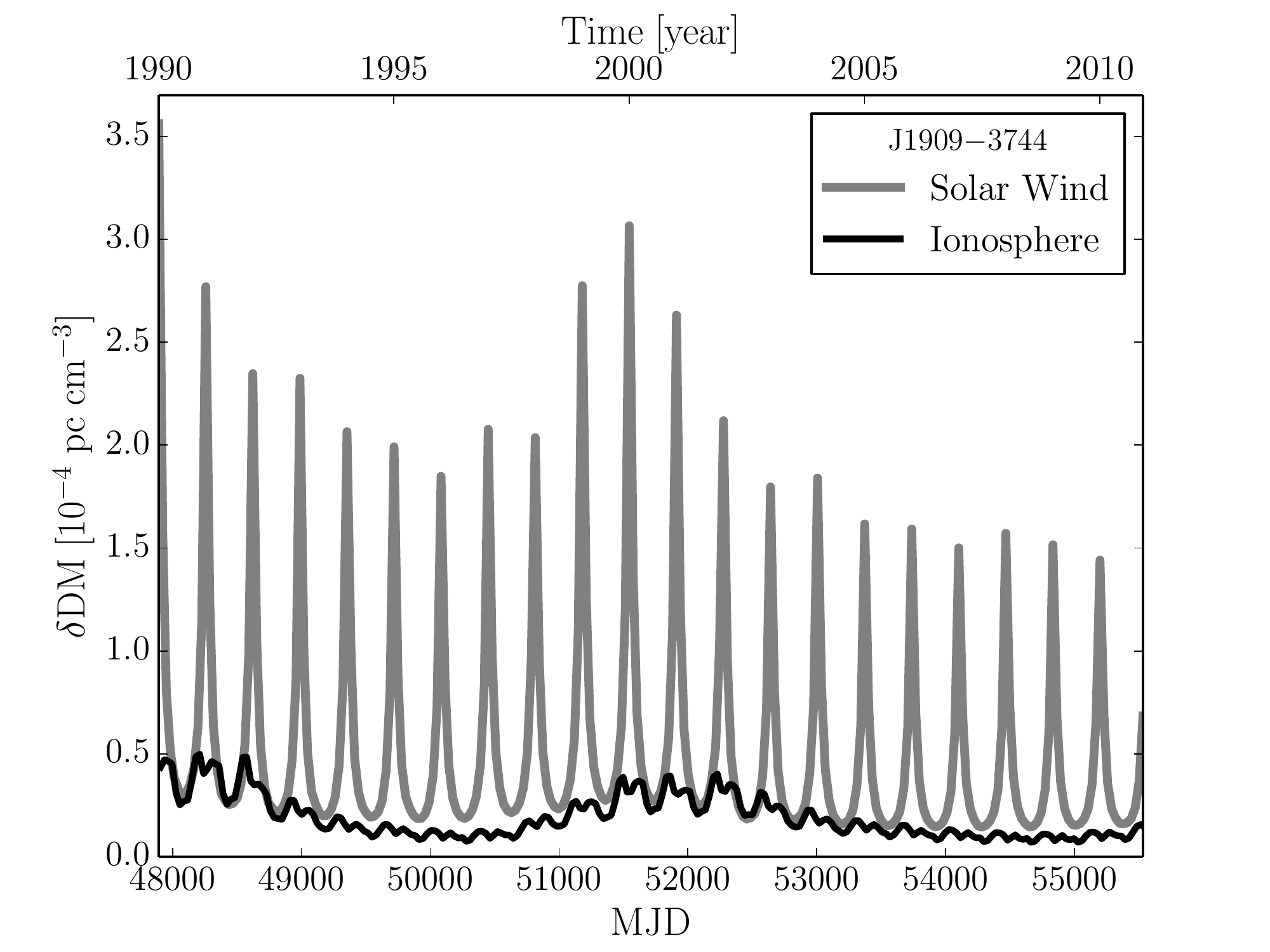}
\caption{Comparison between solar wind and ionospheric DM variations, the latter as computed by the NeQuick 2 model.
\label{fig:comparison}
}
\end{center}
\end{figure}

Figure~\ref{fig:J1909ionosphere} shows our daily estimates of $\DM_\mathrm{ion}(t)$ along the LOS to J1909$-$3744 from both the Green Bank Telescope (GBT, black) in the Northern Hemisphere and the Parkes Telescope (PKS, blue) in the Southern Hemisphere using the method in Eq.~\ref{eq:DMionosphere2}. The top panel shows DM measurements at transit on each day. Error bars come from the GNSS measurement errors alone, multiplied by the same geometric factor dependent on the zenith angle. The overall amplitude shift is a result of the constant difference in the zenith angle of the pulsar at transit between the two sites. Differences in DM between GBT and PKS are shown in the sub-panel beneath. The right panel shows the estimated DM observed with PKS versus GBT, with the solid gray line representing the ratio of the geometric factors. For reference, the median DM measurement error at the GBT for this pulsar is $\sim 1 \times 10^{-5}$ pc \cmthree for the latest backends (see NG9 for information on the GUPPI backend) whereas the value at PKS is $\sim 3\times 10^{-5}$ pc \cmthree. Variations in the DM annually and over the solar cycle are visible. The bottom panels of Figure~\ref{fig:J1909ionosphere} shows the result of simultaneous observations of the pulsar at low elevation angles at both telescopes, such that the ratio of geometric factors is 1:1, i.e., variability is due solely to differences in the ionosphere at different local times. Such an observation occurs 110 minutes after the pulsar transits the GBT when the zenith angle is $\approx 80.3^\circ$, yielding a geometric factor of $\approx 3$ increase over the zenith TEC measurement. The sites are separated by $\sim 112^\circ$ of longitude and observe the pulsar at nearly opposite local times. As the Earth's orbital position shifts, the local observing times shift, producing a phase difference between the two time series. Smaller peaks separated from the yearly peaks by approximately six months are visible as the solar cycle maximum is approached just past the end of the time series \citep[again see][for this intra-annual variability]{hr2006}. Again, the right panel shows the PKS-estimated ionospheric DM versus the GBT-estimated DM, where the estimates in DM can differ by measurable amounts even when observed at the same time. The bimodality results from ionospheric differences between day and night between the sites.

The second method to study the ionospheric DM variations uses the mathematical description above, implemented in the NeQuick 2 model \citep{ncr2008}. NeQuick 2 uses ionosonde measurements to determine the parameters of the semi-Epstein layers (Eq.~\ref{eq:epstein_layer}), along with solar radio flux measurements and a model for the magnetic inclination that describes the shape of the Earth's magnetic field lines as a function of latitude, longitude, and time. For more details on the implementation, refer to \citet{ITUR}. By default, NeQuick 2 contains parameters with a monthly time resolution, and a time series for J1909$-$3744 is shown in comparison to the DM contribution from the solar wind, discussed in the following sub-section, in Figure~\ref{fig:comparison}.

\subsection{Solar Wind}

Particles from the solar corona have enough kinetic energy to escape the Sun's gravity, becoming part of the interplanetary medium. The speeds and compositions of these particles are not uniform, and measurements of the electron density are carried out both from ground-based observing and in situ. \citet{sns+2005} model the electron density along the LOS to PSR J1713+0747 due to the solar wind as a power-law $n_e(r) = n_0~(1~\AU/r)^2$ \cmthree, where $n_0$ is the electron density in \cmthree at the Earth, based on measurements from the {\em Ulysses} spacecraft \citep{Issautier+2001}. They note that while the scaling holds over a large range of heliocentric latitudes, it does not consider spatial variations with ecliptic latitude $\beta_e$, namely the higher-density slow wind at lower latitudes and the lower-density fast wind at higher latitudes, nor does it consider temporal variations. They find that $n_0 = 5\pm4$ \cmthree. \citet{y+07b} present a generic two-piece model that accounts for the positional variations using daily solar magnetic field maps from Wilcox Solar Observatory but do not consider temporal variations; the coefficients for their power-law components come mostly from observations taken at minima in the solar cycle. See references cited by \citet{y+07b} therein for more details.

\citet{sbt+2013} \citep[see also][]{poi+2014} find that the total number density of solar wind protons is $\sim 2-10$ \cmthree at 1~AU over a range of heliolatitudes $\beta_h$ and over the course of a solar cycle. The highest densities comes from $|\beta_h| \lesssim 20^\circ$. At solar cycle maxima, the total density at 1 AU is a weak function of heliolatitiude with a value of $\sim 6$ \cmthree. Heading towards solar cycle minima, the proton density becomes more peaked at central heliolatitudes though the overall quantity drops. We follow the methods in \citet{sbt+2013} to create an empirical model of Carrington rotation-averaged (one period is 27.2753 days) proton density as a function of heliolatitude and time spanning from 1990 to 2011. We linearly interpolate in both heliolatitude and time for smoother sampling of the proton density. Inspection of time series from the {\it Solar Wind Observations Over the Poles of the Sun} (SWOOPS\footnotemark[1]) experiment on the {\it Ulysses} spacecraft show that $n_e \sim n_p$, which we will assume to obtain the electron density at 1~AU, $n_0(\beta_h,t)$ \citep{swoops}. The model assumes an $n_e \propto r^{-2}$ dependence, which is supported elsewhere in the literature \citep[e.g.][]{Issautier+1998}. Therefore, we can write the solar wind DM in the direction of the pulsar as

\footnotetext[1]{\url{ftp://nssdcftp.gsfc.nasa.gov/spacecraft_data/ulysses/plasma/swoops/ion/hires/}}

\noindent
\be
\DM_{\mathrm{sw}}(t,\beta_h) &= &4.848\times 10^{-6} \mathrm{\;pc\;\cmthree} \times \nonumber \\
& & \int \left(\frac{n_0(\beta_h,t)}{\cmthree}\right)\left(\frac{1~\AU}{r}\right)^2 ds,
\label{eq:solar_wind}
\ee
where the integration path $s = s(\beta_h,r)$. We limit the integration to within 100~AU of the Sun. The typical solar wind speed is of the order several hundred kilometers per second and so the propagation time to the integration boundary is of order one year \citep{sbt+2013}. However, because of the $r^{-2}$ factor, only the electron density within the inner $\lesssim 10$~AU contribute to any currently measurable DM variation, which has a propagation time of approximately one Carrington rotation. Since the intrinsic time-averaging with the model is of this order, exclusion of the time-varying mean speeds of the electrons should not greatly affect our results. We note that Eq.~\ref{eq:solar_wind} only accounts for the average behavior of the solar wind over Carrington rotations and does not include components from transient events such as solar flares or coronal mass ejections.

\begin{figure}[t!]
\begin{center}
\includegraphics[width=0.52\textwidth]{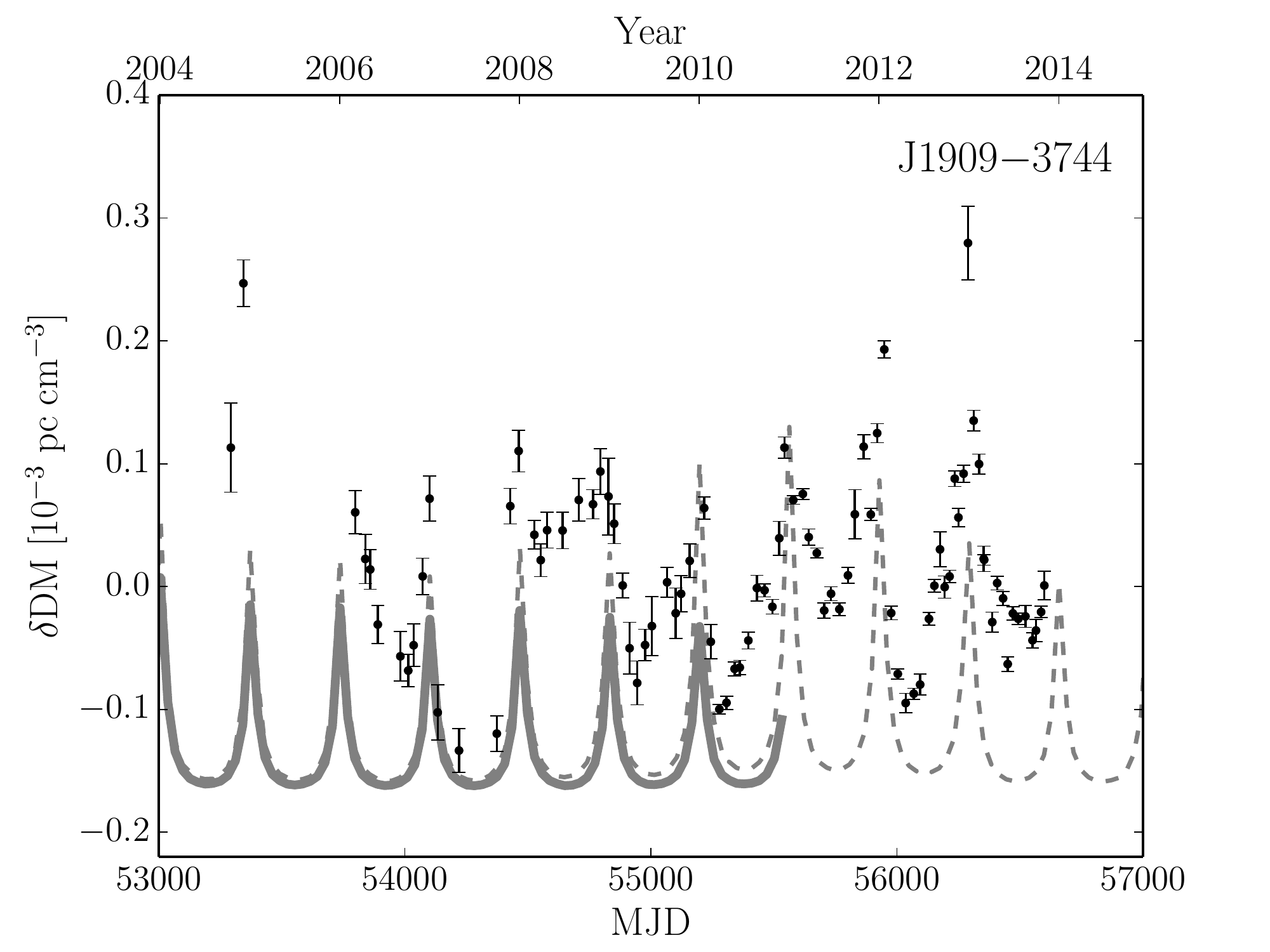}
\caption{Time series of DM for J1909$-$3744 from NG9 after a weighted, quadratic trend has been removed as in Figure~\ref{fig:smallplot}. The solid gray line shows the model \DMsw(t) from Figure~\ref{fig:comparison} with an arbitrary vertical offset added. The dashed gray line shows the model shifted forward in time by 11 years (one average solar cycle) to provide a comparison of the periodicity and shape of the model time series with the data at later times.
\label{fig:solarwind_DMt}
}
\end{center}
\end{figure}

\begin{figure}[t!]
\begin{center}
\includegraphics[width=0.52\textwidth]{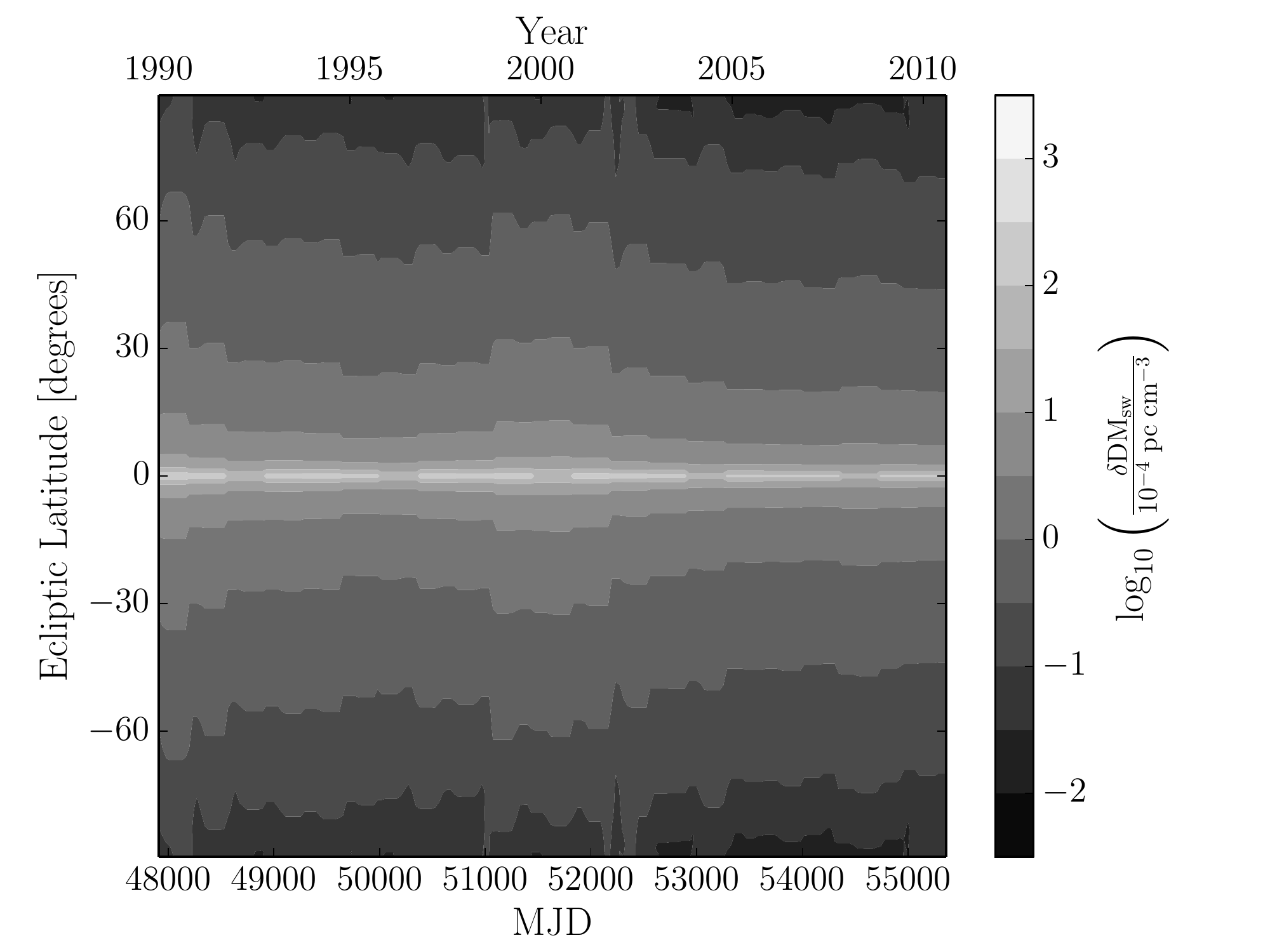}
\caption{Maximum change in model solar wind DM centered around a one-year smoothing window at each given epoch for a pulsar at a given ecliptic latitude.
\label{fig:solarwind_grid}
}
\end{center}
\end{figure}

Figure~\ref{fig:comparison} shows the model solar wind DM and the ionospheric DM from the NeQuick 2 model along the LOS to J1909$-$3744. For the ionospheric component, we set the observation during pulsar transit once per month. Again, the median error on DM for J1909$-$3744 measured with GBT is $\sim 1 \times10^{-5}$ pc \cmthree, which implies that the ionospheric contribution is marginally detectable in the time series, whereas the solar wind contribution is significantly measurable. Figure~\ref{fig:solarwind_DMt} shows the predicted solar wind contribution plotted against the J1909$-$3744 time series with the best-fit quadratic trend removed for clarity. The vertical offset of the predicted time series was set arbitrarily to roughly match the DM offsets (as the nominal DM has already been removed). Even without including transient solar events, our model agrees with the overall periodic trend in the time series, both in phase and peakedness of the yearly maxima. 

Figure~\ref{fig:solarwind_grid} shows the peak-to-peak change in the DM contribution from the solar wind model in a given one-year smoothing window as a function of ecliptic latitude. Pulsars lying closer to the ecliptic plane will have a much greater peak DM since the LOS will cross near the Sun and the electron density scales as $r^{-2}$. Pulsars observed far out of the plane will show minimal amounts of solar wind DM variations. For reference, J1909$-$3744 has $\beta_e \approx -15.2^\circ$ with the mean peak-to-peak change around $1.45 \times 10^{-4}$ pc \cmthree, which can also be seen as the predicted amplitude of variations in Figure~\ref{fig:solarwind_DMt}.

\subsection{Heliosphere}

Particles comprising the solar wind interact with the surrounding ISM at the heliospheric boundary. As the Sun moves through the ISM, it creates a bow shock towards the nose (upwind) direction with a long tail opposite the direction of the Solar System's motion. Turbulence generated at the interface creates spatial and temporal variations in electron density. In general, the DM for a specific LOS can be written as 
\be
\DMhel(t,\beta_e,\lambda_e)\! =\!\!\int\! n_{e,\mathrm{hel}}(s(t,\beta_e,\lambda_e),t) ds
\label{eq:heliosphere}
\ee
where $(\lambda_e,\beta_e)$ are ecliptic longitude and latitude. The path $s$ depends on the position of the Earth in its orbit. For reference, the nose direction of the heliosphere is roughly $(\lambda_e,\beta_e) \approx (254^\circ,5^\circ)$ \citep{kg2003}, equivalent to $(\alpha_e,\delta_e) \approx (253.3^\circ,-17.5^\circ)$ or $(l_g,b_g) \approx (2.6^\circ,16.4^\circ)$.

\begin{figure}[t!]
\hspace{-0.25in}
\includegraphics[width=0.575\textwidth]{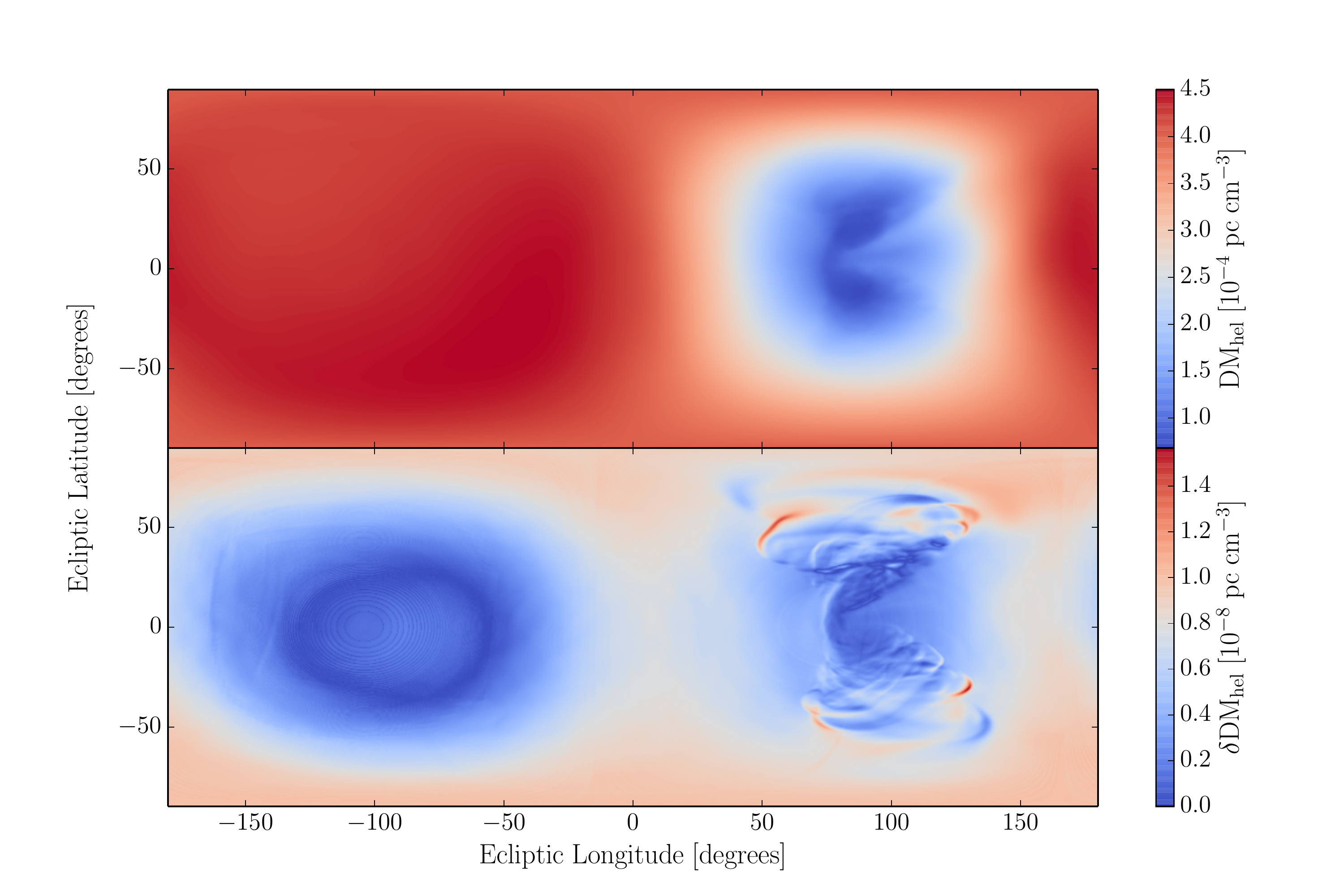} 
\caption{\footnotesize Top: Map of $\DMhel(\lambda_e,\beta_e)$. The nose direction is in the center of the left half of the image whereas the tail direction is in the center of the right half. Bottom: Maximum change in DM due to the Earth's orbital motion around the Sun. Note the different scales between the two panels. The thin, ringed structures visible in the nose direction (not the broad ringed structure) are a result of sampling errors in the 3D grid.
\label{fig:heliosphere_grid}
}
\vspace{2ex}
\end{figure}

\begin{figure*}[t!]
\epsscale{1.2}
\begin{center}
\plotone{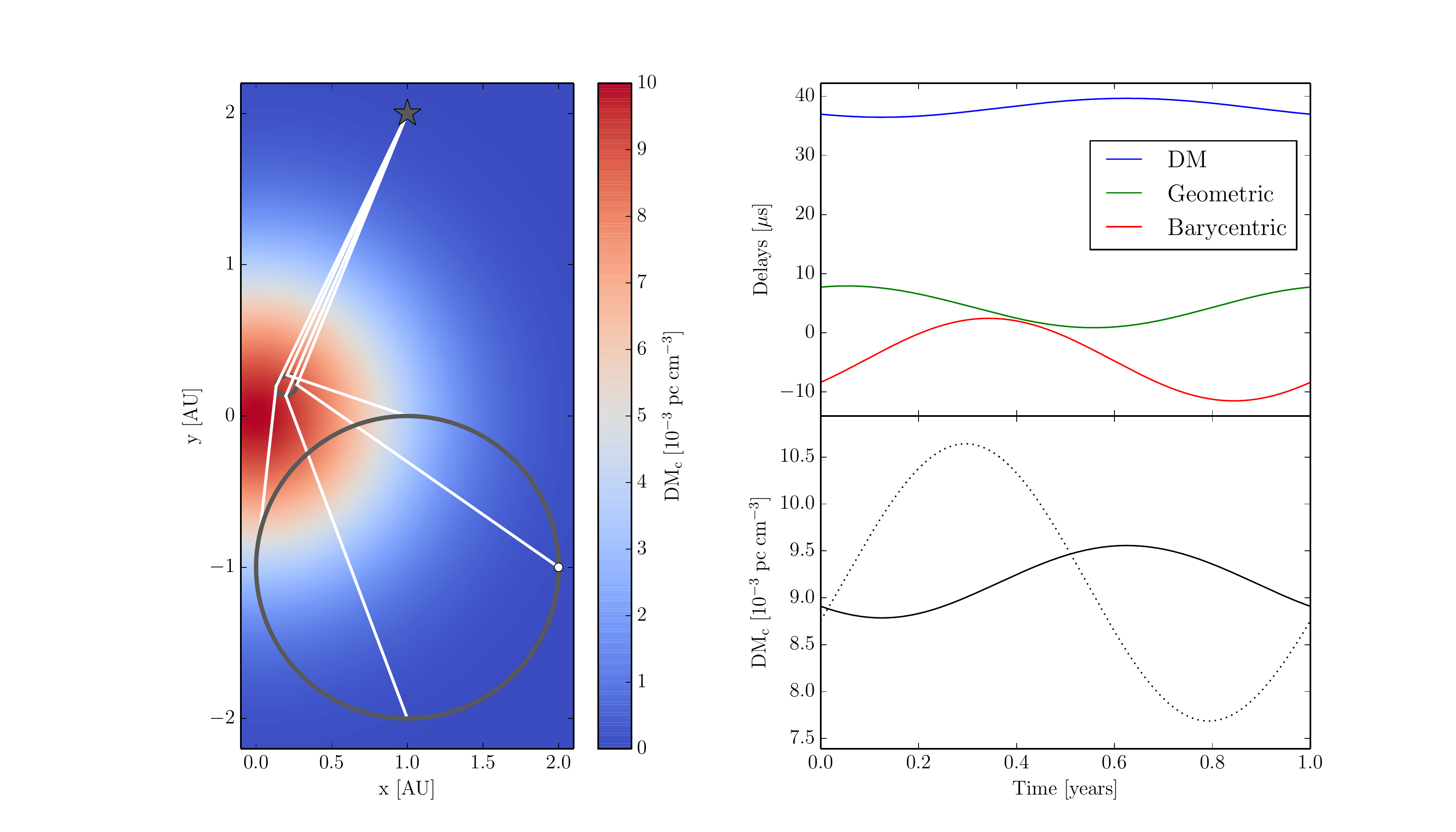}
\caption{Effects of a Gaussian cloud on timing measurements. Left: Example ray-tracing geometry of emission (where the $z$-direction has been collapsed in the image) from a pulsar (gray star) located at (1 AU, 2 AU, 1 kpc) traveling through a cloud with $N_0 = 0.01$~pc~\cmthree and $a=1$~AU located at (0 AU, 0 AU, 0.5 kpc) hitting the Earth in an orbit centered around the Sun located at (1 AU, --1 AU, 0 kpc). The orbit of the Earth (large gray circle, $t=0$~yr is given by the white dot) is in the plane of the image (i.e., $z=0$~AU at all times). The background colors represent the column density profile of the cloud. The smaller gray circle drawn over the cloud represents the sampling of the cloud screen due to the Earth's orbital motion and the position of the pulsar. We integrate at points over the entire orbit; four rays (white lines) have been shown to demonstrate the integration paths from the pulsar to the Earth. Top right: The three delays $\dtDM$ (blue), $\dtgeo$ (green), $\dtbary$ (red) in order from top to bottom. Bottom right: DM due solely to the integral of electron density over the ray paths (solid) as compared with the measured DM when all three delays are summed together (dotted).
\label{fig:gaussianblob}
}
\end{center}
\end{figure*}

\citet{odz+2015} simulate the heliosphere region extending from 30 to 1500~AU. They assume a spherically symmetric solar wind flow at the inner boundary with a given speed, number density, and temperature, along with a radial and azimuthal solar magnetic field. The outer boundary interacts with the ISM and also has a relative velocity, number density, and temperature. The interstellar magnetic field is slanted with respect to the downwind direction. They find that the solar magnetic field forces the solar wind plasma into jets which are then blown into the tail direction by the interstellar wind. Turbulent instabilities form into two tails and the heliosphere retains a two-lobed structure as the tails remain separated. 

Figure~\ref{fig:heliosphere_grid} shows the electron density from simulations in \citet{odz+2015} integrated out to a distance of 1500~AU from the Sun. The heliosphere is several times denser through the bow shock region in the nose direction than through the tail direction. The bottom panel shows the peak-to-peak variations in DM due to the changing LOS from the Earth orbiting the Sun to the stationary pulsar. The maximum change is $\sim 10^{-8}$ pc \cmthree when looking at turbulence through the tail. The effect of the heliosphere, therefore, is approximately three orders of magnitude smaller than current sensitivity (for J1909$-$3744 in NG9) and requires $\sim 0.05$ ns timing precision to measure. While the heliosphere does change over time, the overall structure remains similar. Since the crossing time for solar wind particles through the heliosphere is of the order years, changes in the electron density along a given LOS will be small from epoch to epoch. Given that the overall amplitude of the heliospheric DM is below current sensitivity to DM, we do not consider temporal variations in the heliosphere.

\begin{figure}
\vspace{-2ex}
\begin{center}
\includegraphics[width=0.52\textwidth]{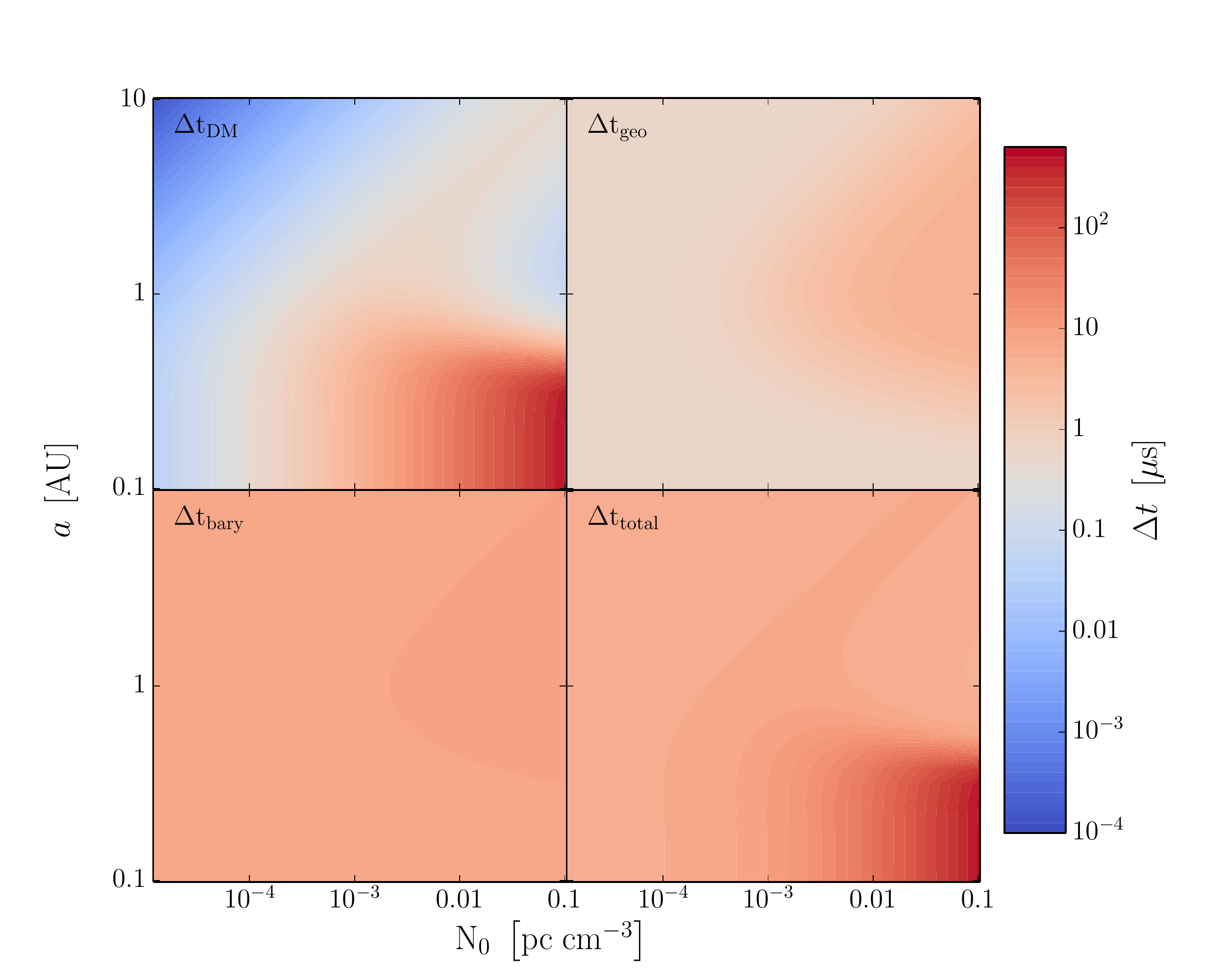}
\caption{Timing delays associated with a refraction due to a Gaussian plasma lens as a function of $N_0$ and $a$. Top left: Dispersion delay $\dtDM$. Top right: Geometric delay $\dtgeo$. Bottom left: Barycentric delay $\dtbary$. Bottom right: Sum of all three delays. The amplitude of the timing delays typically increases towards the bottom right.
\label{fig:gaussianblob_phase_space}
}
\end{center}
\end{figure}

\subsection{Gaussian Plasma Lens in the ISM}

Using the formalism in \S\ref{sec:refraction_effects}, we simulate a Gaussian, electron density cloud and solve Eq.~\ref{eq:ray_equation} to trace incident rays back from the Earth to the pulsar. We show an example calculation of the ray paths through a cloud with central column density $N_0 = 0.01$ pc \cmthree and size $a = 1$ AU in Figure~\ref{fig:gaussianblob}. The pulsar is at a distance of 1 kpc and the cloud is halfway in between.

We numerically integrate through our example cloud and show the delays in the top right panel of Figure~\ref{fig:gaussianblob}. The bottom right panel shows the DM delay purely as the integral of the electron density along the ray path (i.e., proportional to $\dtDM$ alone; solid line) along with the estimated DM when all three delays are summed together (dotted line). In our example, both the amplitude and phase change significantly.

To explore the possible parameter space of $N_0$ and $a$, we place a pulsar at (1 AU, 2 AU, 1 kpc), directly in line with the Gaussian cloud at (0 AU, 0 AU, 0.5 kpc). The Sun is located at (1 AU, --1 AU, 0 kpc) with the Earth orbiting in the $z=0$ plane. Figure~\ref{fig:gaussianblob_phase_space} shows the peak-to-peak variations in the time delays separately and then when all three are added together.

While we consider the case of the LOS crossing through a Gaussian cloud periodically, the same formalism can be applied to single crossing events \citep{1998ApJ...496..253C}. Clouds with a small perpendicular velocity will cause periodic DM variations modulated by an envelope with width equal to the timescale of the cloud crossing. In general, the phase of $\DM_c(t)$ can be arbitrary with respect to previously mentioned periodic contributions.

\section{Implications for ISM Study and Precision Timing}
\label{sec:implications}

Analysis of DM variations can enable the study of the electron density along the entire LOS to a pulsar. While we see that the ISM is consistent with a Kolmogorov medium, interpretations of DM SFs could be used to search for a different wavenumber spectral index or anisotropies along certain LOSs. Measurements of DMs coming from an array of pulsars distributed across the sky has the potential to probe the structure of the ISM and solar wind. Changing DMs due to pulsar motion through the ISM can also be valuable inputs to large-scale electron density models of the Galaxy.

Assuming that the only chromatic effect on pulses is the dispersive delay, DM can be estimated on a per-epoch basis using a wide range in frequency coverage. In that case, variations in DM will not affect pulsar TOAs used for precision timing experiments. Numerous chromatic effects are however known to exist.

\vspace{-0.1in}
\begin{itemize}
\itemsep -1pt
\item Frequency-dependent variations of the pulse profile will change the measured TOAs by a constant offset per frequency and lead to a large error in DM if not globally fit for over a many-epoch data set \citep{Liu+2012,Pennucci+2014}. Profile evolution is assumed to be time-independent in many pulsars (see \citealt{Lyne2010} for counterexamples). A simultaneous fit over parameters that describe the profile evolution and the DM will reduce their covariance. Profile evolution coupled with amplitude modulation from interstellar scintillation will cause an effective shift in the reference frequency that changes the estimated DM on the order of a diffractive timescale.
\item Estimates of DM will be contaminated by other chromatic timing effects that result from refraction and multipath propagation.   As shown by \citet{fc90}, if refraction is allowed to contaminate DM estimates, the SF will show excess amplitude on long times compared to extrapolation from the diffraction timescale and will also lead to an overestimated wavenumber spectral index.  Multipath propagation causes temporal broadening of pulse shapes that increase with lower frequency. Pulse broadening will couple with intrinsic profile shape changes, adding additional time-dependent TOA errors, and thus producing apparent DM changes \citep{2016ApJ...818..166L}.
\item  Scattering causes spatial averaging over the ISM in any single-epoch measurement of DM \citep{css2015}. DM is therefore a function of frequency.
\item DM measured with asynchronous multi-frequency measurements will be mis-estimated as the LOS integral will change, due to stochastic changes in the ISM and to systematic effects such as the increasing Earth-pulsar distance \citep{Lam+2015}.
\item Even simultaneous measurements from different locations on the Earth can result in different observed values of DM due to separate LOSs through the ionosphere.
\end{itemize}
High-precision timing experiments require minimization of all possible TOA errors, especially those correlated in time. Therefore, the combination of data from multiple telescopes will require care to avoid contamination from the various achromatic effects listed.

Inclusion of DM terms that describe linear or periodic variations can reduce the number of model parameters in a timing fit but will also be highly covariant with other parameters included in the fit \citep{sns+2005}. Linear terms for DM evolution in a timing model are covariant with pulsar spin and spin-down parameters. In a pulsar timing array experiment to detect and study gravitational waves (GWs), such terms also remove sensitivity to the lowest-frequency GWs. Annual and semi-annual variations will be covariant with astrometric parameters and GWs with the same frequencies.

Removal of frequency-dependent terms other than DM, such as profile evolution parameters or scattering delays, in a timing model will change the absolute DM measured. The absolute differences must be taken into account when combining DM measurements obtained by different methods; TOAs incur additional errors otherwise. In addition, even with the same frequency-dependent terms included, different methods exist for DM estimation and removal from TOAs. \citet{Keith+2013} utilize information regarding the correlations between epochs in their DM determination; \citet{Demorest+2013} do not. Since DM is not independent from epoch to epoch, timing models should account for the correlations between measurements. However, due to the stochastic component of the DM variations, it may be impossible to completely remove per-epoch DM determination from a timing model. Optimal DM estimation and removal strategies are therefore necessary to minimize TOA uncertainties.

\section{Summary and Conclusions}
\label{sec:conclusions}

DM time series show a wide range of correlated variations. We model the possible contributions to DM variations as the sum of systematic and stochastic effects along the LOS through the media between the observatory and the pulsar. Linear trends arise from the average motion of the LOS through the ISM and the full 3D motion of the pulsar should be taken into consideration when studying linear trends in DM time series. Disentangling the effects of changing distance and changing LOS from parallel and transverse motion, respectively, is possible if scintillation parameters (including flux density) are also measured.   The change in distance over a few years will have no effect on these parameters whereas transverse gradients in the ISM density will.

Changes in the LOS due to Earth's annual motion, coupled with a variety of effects that will be weighted differently for different pulsars, result in periodic variations in DM. Any DM contribution from the ionosphere, solar wind, or heliosphere will be correlated across pulsars depending on their sky positions and the relative position of the Sun. The relative phases of the three contributions may be misaligned, again depending on the specific positions of the pulsars, and so it is possible to disentangle the effects for a subset of pulsars. In the case of the ionosphere, the periodicity may be semi-annual. In general, both types of variations, linear and periodic, will contrbute to DM time series, along with a stochastic component resulting from the turbulent ISM. The relative importance of each component can only be determined on a pulsar-by-pulsar basis.

SFs are useful statistics for analyzing DM variations. DM time series will generally include systematic trends along with stochastic variations from density variations on a wide range of scales (e.g., Kolmogorov-like variations). The stochastic term can be contaminated by any systematic trend in the time series, so time series should be de-trended before using $\DM(t)$ to infer the properties of the ISM along the LOS. Estimates of the wavenumber spectral index or the scintillation timescale from the SF should also include realization errors. We show that once the linear trends and realization errors are taken into account, PSRs J1909$-$3744, B1937+21, and B1821$-$24 show time series consistent with a Kolmogorov electron-density wavenumber spectra. PSR B1534+12, with its non-monotonic trends in DM, is inconsistent with a simple Kolmogorov ISM.

Decomposition of DM time series into known, deterministic causes will allow for the study of the local and interstellar electron density. Future studies of DM time series should model known components to further probe the relative contributions of DM fluctuations along the LOS. Pulsars in a pulsar timing array with many LOSs will see correlated DM variations from the ionosphere and solar wind. As we have shown with J1909$-$3744, we can probe the local electron density around a pulsar after careful determination of its radial velocity.

Differences in DM correction methods will become increasingly important in the near future. Optimal correction methods must be implemented for the proper combination of multi-telescope data. By appropriately removing the effects of DM variations from TOAs, we will be able to maximize pulsar timing array sensitivity.

\acknowledgements

Work on pulsar timing at Cornell University and West Virginia University is supported in part by NSF PIRE program award number 0968296 and NSF Physics Frontier Center award number 1430284. For JWA the research described here was carried out at the Jet Propulsion Laboratory, California Institute of Technology, under a contract with the National Aeronautics and Space Administration. MTL was partially supported by NASA New York Space Grant award number NNX15AK07H. We would like to thank Merav Opher and Bertalan Zieger of Boston University for access to their simulated heliosphere data. We thank Paul Demorest and Ryan Shannon for data and assistance relating to PSR B1937+21, Emmanuel Fonseca for data relating to PSR B1534+12, and Michael Keith for data relating to PSRs J1909$-$3744, B1937+21, and B1821$-$24. We would also like to thank Bruno Nava for the source code to the NeQuick 2 model, David Nice for assistance regarding solar wind modeling, John Antoniadis for discussions on the 3D velocity of PSR J1909$-$3744, the NANOGrav Interstellar Medium Mitigation working group for useful conversations throughout the duration of this project, and the anonymous referee for a very careful readthrough of this work. We acknowledge NASA/GSFC's Space Physics Data Facility's ftp service for {\it Ulysses}/SWOOPS (\url{ftp://nssdcftp.gsfc.nasa.gov/spacecraft_data/ulysses/plasma/swoops/ion/hires/}) data collection.

\appendix

\section{Functional forms for Structure Functions of a Power-Law Wavenumber Spectrum}
\label{appendix:sfs}

We will consider the relationships between different SFs of a time-varying $\DM(t)$. By taking the Fourier transform of the first-order increment $\Delta^{(1)}\DM(t,\tau) = \DM(t) - \DM(t+\tau)$, we can write the ensemble-average SF in terms of the power spectrum $S_\DM(f)$ \citep[see Eq.~15 of][]{Lam+2015},
\be
D^{(1)}_\DM(\tau) &=& \left<\left[\Delta^{(1)} \DM(t,\tau)\right]^2\right> = 4 \int df S_\DM(f) \sin^2(\pi f \tau).
\label{eq:D1_DM}
\ee
A wavenumber spectrum (Eq.~\ref{eq:pne}) with spectral index $\beta$ will be a red noise process $\DM(t)$ with an associated power-law spectrum that scales as $S_\DM(f) = Af^{-\gamma}$ where $\gamma = \beta - 1$ and $A$ is a spectral coefficient. For a wavenumber spectrum in the scintillation regime ($2<\beta<4$, $1<\gamma<3$), the integral can be solved as
\be
4 A \int_0^\infty df f^{-\gamma} \sin^2(\pi f \tau) = -2A\Gamma\left(-[\gamma-1]\right)\cos\left(\frac{\pi[\gamma-1]}{2}\right)(2\pi\tau)^{\gamma-1},
\ee
where $\Gamma$ is the Gamma function \citep[][Eq.~3.823]{GR}. We can relate the spectral coefficient $A$ to the scintillation timescale $\Dtiss$ by equating this to Eq.~\ref{eq:D_DM_tau_dtiss},
\be
A = -\frac{\nu^2}{2(cr_e)^2 \Gamma(-[\gamma-1])\cos\left(\pi[\gamma-1]/2\right)\left(2\pi\Dtiss\right)^{\gamma-1}}.
\label{eq:A_to_dtiss}
\ee
Therefore, a Kolmogorov wavenumber spectrum with $\beta = 11/3$ and a time-series power-law spectral index of $\gamma = 8/3$ will have a SF equal to
\be
D^{(1)}_\DM(\tau) = A\Gamma\left(-5/3\right)\sqrt{3}(2\pi\tau)^{5/3}
\label{eq:D1_DM_tau}
\ee
with
\be
A = -\frac{\nu^2}{(cr_e)^2 \Gamma(-5/3)\sqrt{3}\left(2\pi\Dtiss\right)^{5/3}},
\ee
which reduces to Eq.~\ref{eq:D_DM_kolmogorov} when combined. In general, the scintillation timescale will vary with frequency as $\Dtiss \propto \nu^{2/(\beta-2)}$, so the SF, proportional to $\nu^2 [\Dtiss(\nu)]^{-(\beta-2)}$, will always be independent of frequency in the scintillation regime ($2<\beta<4$) \citep{Lam+2015}.

Following a similar procedure using the second-order increment  $\Delta^{(2)}\DM(t,\tau) = \DM(t-\tau) - 2\DM(t) + \DM(t+\tau)$, the second-order SF can be written as  \citep[Eq.~21 of][]{Lam+2015}
\be
D^{(2)}_\DM(\tau) &=& \left<\left[\Delta^{(2)} \DM(t,\tau)\right]^2\right> = 16 \int df S_\DM(f) \sin^4(\pi f \tau)
\label{eq:D2_DM}
\ee
Using trigonometic identities, we can write $\sin^4(\theta) =  \sin^2(\theta) - (1/4)\sin^2(2\theta)$ and then solve using similar integrals to before. The second-order SF can then be related to the first-order SF for a power-law spectrum
\be
D^{(2)}_\DM(\tau) & = & -\left(8-2^{-\gamma}\right)A\Gamma\left(-(\gamma-1)\right)\cos\left(\frac{\pi(\gamma-1)}{2}\right)(2\pi\tau)^{\gamma-1} = \left(4-2^{\gamma-1}\right)D^{(1)}_\DM(\tau),
\ee
which is roughly $0.8252 D^{(1)}_\DM(\tau)$ for the Kolmogorov case. This is equal to the variance of the second-order increments, $\sigma^2_{\Delta^{(2)}\DM}(\tau)$. Thus, for a pulsar with scintillation timescale $\Dtiss(\nu)$ measured at frequency $\nu$, the second-order increments at a given $\tau$ will be drawn from a Gaussian distribution with standard deviation
\be
\sigma_{\Delta^{(2)}\DM}(\tau) = \left[D^{(2)}_\DM(\tau)\right]^{1/2} = \sqrt{4-2^{\gamma-1}}\left[D^{(1)}_\DM(\tau)\right]^{1/2} = \sqrt{4-2^{\beta-2}}\left(\frac{\nu}{c r_e}\right) \left[\frac{\tau}{\Dtiss(\nu)}\right]^{(\beta-2)/2}
\ee
While Eq.~\ref{eq:sigma_DM} has the variance of $\DM(t)$ equal to half the DM SF, we note that the variance of DM increments will be equal to the SF only, which is defined as the expectation value of the square of the increments.

\section{Structure Function Slope Mis-Estimation from Additive Noise}
\label{appendix:sf_error}

The presence of additive noise will also bias the slope $\alpha$ of a power-law SF, $D_x(\tau) = C \tau^\alpha$, for a time series $x(t)$. Assuming for now that a linear trend has been removed, since the SFs listed above all have the same slope for a Kolmogorov medium, we let $y(t) = x(t) + n(t)$ be the measured values of a generic, random process, where $x(t)$ is the random process of interest (e.g., DM variations) and $n(t)$ is the measurement error with rms $\sigma_n$. The SF of $y(t)$ is then
\be
D_y(\tau) = \left(1-\delta_{\tau 0}\right)2\sigma_n^2 + D_x(\tau)
\label{eq:DMhatSF}
\ee
where $\delta_{\tau 0}$ is the Kronecker delta. The slope of the SF of $y$ might be used as an estimate for $\alpha$. For $\tau > 0$, it can be shown that the estimated slope is
\be
\hat{\alpha} = \frac{d\ln D_y(\tau)}{d\ln\tau} \equiv \frac{\tau}{D_y(\tau)}\frac{dD_y(\tau)}{d\tau} = \alpha \left[\frac{\tau C\tau^{\alpha-1}}{D_y(\tau)}\right] = \frac{\alpha}{1+2\sigma_n^2 / D_x(\tau)}.
\ee
Therefore, we see that $\alpha$ is always underestimated if the additive noise contribution to the SF is significant. One method of mitigatng the bias is to use a model for the SF that includes a constant term, $\hat{D}_y(\tau) = c\tau^a + b$, where estimates of the three parameters of the least-squares fit would correspond to $\alpha$, $2\sigma_n^2$, and $C$. It is better to do the fit in log-log space because the dynamic ranges of $\tau$ and $D_y(\tau)$ can be large. The estimated slope $\hat{\alpha}$ will take a more complicated form if the linear trend has not been removed.


\vspace{5ex}


\begin{thebibliography}{27}
\expandafter\ifx\csname natexlab\endcsname\relax\def\natexlab#1{#1}\fi


\bibitem[Antoniadis(2013)]{Antoniadis} Antoniadis, J.~I.\ 2013, Ph.D.~thesis, Univ. of Bonn

\bibitem[Armstrong et al.(1995)]{ars1995} Armstrong, J.~W., Rickett, B.~J., \& Spangler, S.~R.\ 1995, \apj, 443, 209 

\bibitem[Arzoumanian et al.(2015a)]{NG9BWM} Arzoumanian, Z., Brazier, A., Burke-Spolaor, S., et al.\ 2015, \apj, 810, 150 


\bibitem[Arzoumanian et al.(2015b)]{Arzoumanian+2015} Arzoumanian, Z., Brazier, A., Burke-Spolaor, S., et al.\ 2015, \apj, 813, 65

\bibitem[Backer et al.(1993)]{b+93} Backer, D.~C., Hama, S., van Hook, S., \& Foster, R.~S.~1993, \apj, 404, 636



\bibitem[Bame et al.(1992)]{swoops} Bame, S.~J., McComas, D.~J., Barraclough, B.~L., et al.\ 1992, \aaps, 92, 237 



\bibitem[Bogdanov et al.(2002)]{Bogdanov+2002} Bogdanov, S., Pruszy{\'n}ska, M., Lewandowski, W., \& Wolszczan, A.\ 2002, \apj, 581, 495 

\bibitem[Brisken et al.(2010)]{Brisken+2010} Brisken, W.~F., Macquart, J.-P., Gao, J.~J., et al.\ 2010, \apj, 708, 232 

\bibitem[Brownsberger \& Romani(2014)]{br2014} Brownsberger, S., \& Romani, R.~W.\ 2014, \apj, 784, 154 


\bibitem[Chatterjee \& Cordes(2004)]{cc2004} Chatterjee, S., \& Cordes, J.~M.\ 2004, \apjl, 600, L51 

\bibitem[Clegg et al.(1988)]{Clegg+1988} Clegg, A.~W., Chernoff, 
D.~F., \& Cordes, J.~M.\ 1988, Radio Wave Scattering in the Interstellar Medium, 174, 174 

\bibitem[Clegg et al.(1998)]{1998ApJ...496..253C} Clegg, A.~W., Fey, A.~L., 
\& Lazio, T.~J.~W.\ 1998, \apj, 496, 253 

\bibitem[Cognard 
\& Lestrade(1997)]{1997A&A...323..211C} Cognard, I., \& Lestrade, J.-F.\ 1997, \aap, 323, 211 

\bibitem[Coles et al.(1987)]{Coles+1987} Coles, W.~A., Rickett, B.~J., Codona, J.~L., \& Frehlich, R.~G.\ 1987, \apj, 315, 666

\bibitem[Coles et al.(2015)]{Coles+2015} Coles, W.~A., Kerr, M., Shannon, R.~M., et al.\ 2015, \apj, 808, 113 

\bibitem[Cordes(2002)]{Cordes2002} Cordes, J.~M.\ 2002, Single-Dish Radio Astronomy: Techniques and Applications, 278, 227

\bibitem[Cordes et al.(1986)]{Cordes+1986} Cordes, J.~M., Pidwerbetsky, A., \& Lovelace, R.~V.~E.\ 1986, \apj, 310, 737 

\bibitem[Cordes et al.(1990)]{1990ApJ...349..245C} Cordes, J.~M.,
Wolszczan, A., Dewey, R.~J., Blaskiewicz, M.,
\& Stinebring, D.~R.\ 1990, \apj, 349, 245

\bibitem[Cordes \& Lazio(1991)]{cl1991} Cordes, J.~M., \& Lazio, T.~J.\ 1991, \apj, 376, 123 

\bibitem[Cordes \& Rickett(1998)]{cr98}
Cordes, J. M. \& Rickett, B. J.\ 1998, \apj, 507, 846

\bibitem[Cordes \& Lazio(2002)]{NE2001} Cordes, J.~M., \& Lazio, T.~J.~W.\ 2002, arXiv:astro-ph/0207156 

\bibitem[Cordes et al.(2015)]{css2015} Cordes, J.~M., Shannon, 
R.~M., \& Stinebring, D.~R.\ 2015, arXiv:1503.08491 

\bibitem[Demorest et al.(2013)]{Demorest+2013} Demorest, P.~B., 
Ferdman, R.~D., Gonzalez, M.~E., et al.\ 2013, \apj, 762, 94

\bibitem[Dow et al.(2009)]{IGS} Dow, J.M., Neilan, R. E., \& Rizos, C.\ 2009, The International GNSS Service in a changing landscape of Global Navigation Satellite Systems, Journal of Geodesy 83:191–198 

\bibitem[Draine(2011)]{Draine} Draine, B.~T.\ 2011, Physics of 
the Interstellar and Intergalactic Medium by Bruce T.~Draine.~Princeton 
University Press, 2011.~ISBN: 978-0-691-12214-4,  

\bibitem[Faucher-Gigu{\`e}re \& Kaspi(2006)]{fk2006} Faucher-Gigu{\`e}re, C.-A., \& Kaspi, V.~M.\ 2006, \apj, 643, 332 


\bibitem[Fonseca et al.(2014)]{fst2014} Fonseca, E., Stairs, 
I.~H., \& Thorsett, S.~E.\ 2014, \apj, 787, 82

\bibitem[Foster \& Cordes(1990)]{fc90}  
Foster, R.~S., \& Cordes, J.~M.\ 1990, \apj, 364, 123

\bibitem[Freire et al.(2001)]{fcl2001} Freire, P.~C., Camilo, F., Lorimer, D.~R., et al.\ 2001, \mnras, 326, 901 

\bibitem[Frisch(2007)]{2007ASPC..365..227F} Frisch, P.~C.\ 2007, SINS - 
Small Ionized and Neutral Structures in the Diffuse Interstellar Medium, 
365, 227

\bibitem[Frisch et 
al.(2011)]{frs2011} Frisch, P.~C., Redfield, S., \& Slavin, J.~D.\ 2011, \araa, 49, 237 

\bibitem[Gaensler et al.(2002)]{gjs2002} Gaensler, B.~M., Jones, D.~H., \& Stappers, B.~W.\ 2002, \apjl, 580, L137 

\bibitem[Gibson(2007)]{2007ASPC..365...59G} Gibson, S.~J.\ 2007, SINS - 
Small Ionized and Neutral Structures in the Diffuse Interstellar Medium, 
365, 59 


\bibitem[Gradshteyn et al.(2007)]{GR} Gradshteyn, I.~S., 
Ryzhik, I.~M., Jeffrey, A., \& Zwillinger, D.\ 2007, Table of Integrals, Series, and Products, Seventh Edition by I.~S.~Gradshteyn, I.~M.~Ryzhik, Alan Jeffrey, and Daniel Zwillinger.~Elsevier Academic Press, 2007.~ISBN 012-373637-4 

\bibitem[Gupta et al.(1994)]{1994MNRAS.269.1035G} Gupta, Y., Rickett, 
B.~J., \& Lyne, A.~G.\ 1994, \mnras, 269, 1035

\bibitem[Hamilton et al.(1985)]{hhc1985} Hamilton, P.~A., Hall, P.~J., \& Costa, M.~E.\ 1985, \mnras, 214, 5P 

\bibitem[Hobbs et al.(2004)]{Hobbs+2004} Hobbs, G., Lyne, A.~G., Kramer, M., Martin, C.~E., \& Jordan, C.\ 2004, \mnras, 353, 1311 

\bibitem[Huang \& Roussel-Dupr\'{e}(2006)]{hr2006} Huang, Z., \& Roussel-Dupr\'{e}, R.\ 2006, Radio Science, 41, RS6004

\bibitem[Ilyasov et al.(2005)]{2005AstL...31...30I} Ilyasov, Y.~P., Imae, 
M., Hanado, Y., et al.\ 2005, Astronomy Letters, 31, 30 

\bibitem[Isaacman \& Rankin(1977)]{ir1977} Isaacman, R., \& Rankin, J.~M.\ 1977, \apj, 214, 214 


\bibitem[Issautier et al.(1998)]{Issautier+1998} Issautier, K., Meyer-Vernet, N., Moncuquet, M., \& Hoang, S.\ 1998, \jgr, 103, 1969 


\bibitem[Issautier et al.(2001)]{Issautier+2001} Issautier, K., Hoang, 
S., Moncuquet, M., \& Meyer-Vernet, N.\ 2001, \ssr, 97, 105 

\bibitem[Report ITU-R P.2297-0(2013)]{ITUR} International Telecommunications Union 2013, Electron Density Models and Data for Transionospheric Radio Propagation, Report ITU-R P.2297-0 (Geneva, Switzerland)


\bibitem[Jacoby et al.(2005)]{Jacoby2005} Jacoby, B.~A., Hotan, A., Bailes, M., Ord, S., \& Kulkarni, S.~R.\ 2005, \apjl, 629, L113 

\bibitem[Kaspi et al.(1994)]{ktr1994} Kaspi, V.~M., Taylor, J.~H., \& Ryba, M.~F.\ 1994, \apj, 428, 713

\bibitem[Keith et al.(2013)]{Keith+2013} Keith, M.~J., Coles, W., 
Shannon, R.~M., et al.\ 2013, \mnras, 429, 2161

\bibitem[Kurth \& Gurnett(2003)]{kg2003} Kurth, W.~S., \& Gurnett, D.~A.\ 2003, Journal of Geophysical Research (Space Physics), 108, 8027 

\bibitem[Lam et al.(2015)]{Lam+2015} Lam, M.~T., Cordes, J.~M., 
Chatterjee, S., \& Dolch, T.\ 2015, \apj, 801, 130 

\bibitem[Levin et al.(2016)]{2016ApJ...818..166L} Levin, L., McLaughlin, M.~A., Jones, G., et al.\ 2016, \apj, 818, 166 

\bibitem[Liu et al.(2012)]{Liu+2012} Liu, K., Keane, E.~F., Lee, K.~J., et al.\ 2012, \mnras, 420, 361 

\bibitem[L{\"o}hmer et al.(2001)]{l+01} L{\"o}hmer, O., 
Kramer, M., Mitra, D., Lorimer, D.~R., 
\& Lyne, A.~G.\ 2001, \apjl, 562, L157 

\bibitem[L{\"o}hmer et al.(2004)]{lo+04} L{\"o}hmer, O., Mitra, D., Gupta, Y., Kramer, M., \& Ahuja, A.\ 2004, \aap, 425, 569 

\bibitem[Lorimer \& Kramer(2012)]{handbook} Lorimer, D.~R., \& Kramer, M.\ 2012, Handbook of Pulsar Astronomy, by D.~R.~Lorimer , M.~Kramer, Cambridge, UK: Cambridge University Press, 2012

\bibitem[Lyne et al.(2010)]{Lyne2010} Lyne, A., Hobbs, G., Kramer, M., Stairs, I., \& Stappers, B.\ 2010, Science, 329, 408 


\bibitem[Maitia et al.(2003)]{2003ApJ...582..972M} Maitia, V., Lestrade, 
J.-F., \& Cognard, I.\ 2003, \apj, 582, 972 


\bibitem[Manchester et al.(1991)]{mlr1991} Manchester, R.~N., Lyne, A.~G., Robinson, C., Bailes, M., \& D'Amico, N.\ 1991, \nat, 352, 219 

\bibitem[Manchester et al.(2005)]{PSRCAT} Manchester, R.~N., Hobbs, G.~B., Teoh, A., \& Hobbs, M.\ 2005, \aj, 129, 1993 

\bibitem[McClure-Griffiths et al.(2007)]{2007ASPC..365...65M} 
McClure-Griffiths, N.~M., Dickey, J.~M., Gaensler, B.~M., Green, A.~J., 
\& Haverkorn, M.\ 2007, SINS - Small Ionized and Neutral Structures in the Diffuse Interstellar Medium, 365, 65

\bibitem[Nava et al.(2008)]{ncr2008} Nava, B., Co{\"i}sson, P., and Radicella, S.~M.\ (2008), Journal of Atmosphere and Solar-Terrestrial Physics, 70(15):1856-1862.

\bibitem[Opher et al.(2015)]{odz+2015} Opher, M., Drake, J.~F., Zieger, B., \& Gombosi, T.~I.\ 2015, \apjl, 800, L28 

\bibitem[Pennucci et al.(2014)]{Pennucci+2014} Pennucci, T.~T., Demorest, P.~B., \& Ransom, S.~M.\ 2014, \apj, 790, 93 

\bibitem[Petroff et al.(2013)]{2013MNRAS.435.1610P} Petroff, E., Keith, M.~J., Johnston, S., van Straten, W., \& Shannon, R.~M.\ 2013, \mnras, 435, 1610

\bibitem[Phillips \& Wolszczan(1991)]{1991ApJ...382L..27P} Phillips, J.~A., \& Wolszczan, A.\ 1991, \apjl, 382, L27

\bibitem[Provornikova et al.(2014)]{poi+2014} Provornikova, E., Opher, M., Izmodenov, V.~V., Richardson, J.~D., \& Toth, G.\ 2014, \apj, 794, 29 

\bibitem[Ramachandran et al.(2006)]{Ramachandran+2006} Ramachandran, R., Demorest, P., Backer, D.~C., Cognard, I., \& Lommen, A.\ 2006, \apj, 645, 303

\bibitem[Rankin \& Roberts(1971)]{rr1971} Rankin, J.~M., \& Roberts, J.~A.\ 1971, The Crab Nebula, 46, 114 

\bibitem[Rawer(1982)]{Rawer1982} Rawer, K.\ Advances in Space Research, 1982, 2, 10, 183

\bibitem[Reardon et al.(2016)]{Reardon+2016} Reardon, D.~J., Hobbs, G., Coles, W., et al.\ 2016, \mnras, 455, 1751 


\bibitem[Rickett(1990)]{r90}    
Rickett, B.~J.\ 1990, \araa, 28, 56

\bibitem[Rickett et al.(2000)]{Rickett+2000} Rickett, B.~J., Coles,  W.~A., \& Markkanen, J.\ 2000, \apj, 533, 304 



\bibitem[Sok{\'o}{\l} et al.(2013)]{sbt+2013} Sok{\'o}{\l}, J.~M., Bzowski, M., Tokumaru, M., Fujiki, K., \& McComas, D.~J.\ 2013, \solphys, 285, 167 

\bibitem[Splaver et al.(2005)]{sns+2005} Splaver, E.~M., Nice, 
D.~J., Stairs, I.~H., Lommen, A.~N., \& Backer, D.~C.\ 2005, \apj, 620, 405 

\bibitem[Stanimirovi{\'c} et al.(2007)]{2007ASPC..365...22S} 
Stanimirovi{\'c}, S., Heiles, C., 
\& Kanekar, N.\ 2007, SINS - Small Ionized and Neutral Structures in the Diffuse Interstellar Medium, 365, 22 

\bibitem[Stinebring et al.(2000)]{Stinebring+2000} Stinebring, D.~R., Smirnova, T.~V., Hankins, T.~H., et al.\ 2000, \apj, 539, 300 

\bibitem[Stinebring(2007)]{2007ASPC..365..254S} Stinebring, D.\ 2007, SINS
- Small Ionized and Neutral Structures in the Diffuse Interstellar Medium, 365, 254

\bibitem[You et al.(2007)]{y+07} 
You, X.~P., Hobbs, G., Coles, W.~A., et al.\ 2007, \mnras, 378, 493 

\bibitem[You et al.(2007)]{y+07b} You, X.~P., Hobbs, G.~B., 
Coles, W.~A., Manchester, R.~N., \& Han, J.~L.\ 2007, \apj, 671, 907

\end{thebibliography}
\end{document}